\documentclass[final,5p,times,twocolumn]{elsarticle}

\usepackage{amssymb}
\usepackage{amsmath}
\usepackage{amsthm}
\usepackage{lineno}
\usepackage{hyperref}
\usepackage{fleqn}
\usepackage{xcolor}
\journal{Chaos, Solitons \& Fractals}

\begin{document}

\begin{frontmatter}

\title{Nonlinear quadrupole topological insulators}

\author[a]{Rujiang~Li\corref{cor1}}
\cortext[cor1]{Corresponding author.}
\ead{rujiangli@xidian.edu.cn}
\author[a]{Wencai~Wang}
\author[a]{Yongtao~Jia}
\author[a]{Ying~Liu}
\author[b,c]{Pengfei~Li}
\author[d,e]{Boris~A.~Malomed}

\affiliation[a]{organization={National Key Laboratory of Radar Detection and Sensing, School of Electronic Engineering, Xidian University},
            city={Xi'an},
            postcode={710071},
            country={China}}

\affiliation[b]{organization={Department of Physics, Taiyuan Normal University},
            city={Jinzhong},
            postcode={030619},
            country={China}}

\affiliation[c]{organization={Institute of Computational and Applied Physics, Taiyuan Normal University},
            city={Jinzhong},
            postcode={030619},
            state={Shanxi},
            country={China}}

\affiliation[d]{organization={Department of Physical Electronics, School of Electrical Engineering, Faculty
of Engineering, Tel Aviv University},
            city={Tel Aviv},
            postcode={69978},
            country={Israel}}

\affiliation[e]{organization={Instituto de Alta Investigaci\'{o}n, Universidad de Tarapac\'{a}},
            addressline={Casilla 7D},
            city={Arica},
            country={Chile}}

\begin{abstract}
Higher-order topological insulators (HOTIs) represent a family of
topological phases that go beyond the conventional bulk-boundary
correspondence. $d$-dimensional $n$-th order HOTIs maintain $\left(
d-n\right) $-dimensional gapless boundary states (in particular,
zero-dimensional corner states in the case of $d=n=2$). HOTIs of the Wannier
type cam be extended into the nonlinear regime. Another prominent class of
HOTIs, in the form of multipole insulators, was investigated only in the
linear regime, due to the challenge of simultaneously achieving both
negative hopping and strong nonlinearity. Here we propose the concept of
nonlinear quadrupole topological insulators (NLQTIs) and report their \emph{%
experimental realization} in an electric circuit lattice. Quench-initiated
dynamics gives rise to nonlinear topological corner states and topologically
trivial corner solitons, in weakly and strongly nonlinear regimes,
respectively. Furthermore, we reveal the formation of two distinct types of
bulk solitons, one existing in the middle finite gap under the action of
weak nonlinearity, and another one found in the semi-infinite gap under
strong nonlinearity. This work realizes another member of the nonlinear
HOTI family, suggesting directions for exploring novel solitons across a
broad range of topological insulators.
\end{abstract}

\begin{keyword}
Quadrupole topological insulators \sep
Nonlinear higher-order topological insulators \sep
Nonlinear corner states \sep Bulk solitons \sep Topological circuits

\end{keyword}

\end{frontmatter}

\section{Introduction}

Standard topological insulators (TIs) maintain topological edge states with
dimension lower by $1$ than that of the bulk \cite{RMP82-3045, RMP83-1057}.
Extending the notion of the electric polarization to higher-order multipole
moments, TIs with a multipole structure have been proposed \cite%
{science357-61, PRB96-245115}. They represent higher-order topological
phases extending the conventional bulk-boundary correspondence. In
particular, a two-dimensional (2D) quadrupole topological insulator (QTI)
features the hallmark of topologically protected zero-dimensional corner
states. This possibility addresses the long-standing issue of whether higher
multipole moments can take quantized values and thus maintain topological
phases, a problem that remained unresolved for a long time. Multipole TIs
have been realized in various physical systems, including mechanical \cite%
{nature555-342}, photonic \cite{nature555-346, nphoton13-692},
electric-circuit \cite{nphys14-925, LSA9-1, PRB100-201406, PRB102-100102},
and acoustic \cite{ncommun11-2108, PRL124-206601, ncommun11-2442,
ncommun11-65} lattices, a crucial step being the implementation of positive
and negative hoppings in them, where the positive hopping corresponds to the
coupling of adjacent lattice sites with no phase shift, while the negative
hopping carries a phase shift of $\pi $ (the sign inversion) \cite%
{nature555-346,PRL124-206601,ncommun11-2442}. These hoppings can be
implemented in staggered optical and matter-wave settings \cite%
{staggered1,staggered2} and in electric-circuit lattices with an appropriate
combination of capacitors and inductors \cite{nphys14-925, LSA9-1,
PRB100-201406, PRB102-100102}. On the other hand, to eliminate the
requirement of the negative hopping, another class of HOTIs, based on
\textquotedblleft quantized" Wannier centers (ones with discrete location of
their centers), was proposed \cite{PRL120-026801, PRB98-045125} and
experimentally demonstrated in a variety of physical setups \cite%
{nmater18-113, nmater18-108, nphoton12-408, nphys15-582, PRL122-233903,
nmat18-1292, PRL122-204301, PRL122-233902, nphoton13-697, sciadv6-eaay4166,
ncommun10-5331,ncommun16-3122}.

There is growing interest in exploring the interplay between topological
states and intrinsic nonlinearity in standard (first-order) TIs \cite{APR7-021306,
nphys20-905,arxiv-lee-1,arxiv-lee-2,PR1093-1} and higher-order TIs (HOTIs)
\cite{ISAN,comment,AM37-2500556,CP8-451}. Nonlinearity not only extends topological
states from the linear limit \cite{PRA90-023813,
PRL117-143901, PRA94-021801, PRL128-093901,ncommun15-9642, PRX11-041057, PRL121-163901,
LSA9-147, PRB102-115411, PRL133-116602, ncommun16-422, ncommun11-1902,
nphys17-1169, science384-1356,arxiv}, but also gives rise to bulk solitons
that do not exist in the linear regime \cite{arxiv,PRL111-243905, science368-856,
optica10-1310, PRL129-135501, nphys18-678, PRL118-023901, LPR13-1900223,
CP5-275, PRB105-L201111}. Building on theoretical studies \cite%
{NJP22-103058, OL45-4710, PRB104-235420, PRA107-033514, CSF185-115188,
PRB110-104307,NJP26-063004}, nonlinear Wannier-type HOTIs have been
experimentally demonstrated, featuring topological phase transitions \cite%
{PRL123-053902}, higher-order topological bound states in the continuum \cite%
{LSA10-164}, and formation of nonlinear topological corner states, along
with topologically trivial corner solitons \cite{nphys17-995}. In contrast,
multipole TIs, while being the earliest proposed and most representative
type of HOTIs, have been addressed only in the framework of the linear
regime, primarily due to the challenges of realizing the negative hopping
and strong nonlinearity simultaneously. QTIs with externally controlled
hoppings\ also remain essentially linear objects, as they lack
amplitude-dependent self-interaction \cite{PRB99-020304}.

The objective of the present work is to introduce the concept of nonlinear
QTIs (NLQTIs) and \emph{experimentally demonstrate} their realization in an
electric-circuit lattice. Similar to the nonlinear Wannier-type HOTIs
realized in evanescently-coupled optical-waveguide lattices \cite%
{nphys17-995}, we initiate dynamical regimes by the application of quench
and observe the excitation of NLQTIs in the form of weakly nonlinear
topological corner states and strongly nonlinear topologically trivial
corner solitons. Additionally, we discover the formation of two distinct
types of bulk solitons, \textit{viz}., weakly and strongly nonlinear ones,
which populate the middle finite bandgap and semi-infinite gap,
respectively. These solitons have not been predicted or observed before in
nonlinear HOTIs (including nonlinear Wannier-type ones). Thus, the present
work extends multipole TIs, in the form of NLQTIs, into the nonlinear regime.
Since only two types of HOTIs have been discovered so far, our NLQTI may be
considered as the another constituent of the nonlinear HOTI family.
Further, our results overcome the previous limitation
which implied that bulk solitons were only found in the standard
(first-order) TIs \cite{science368-856, optica10-1310,nphys18-678}. This
framework can be further extended to include nonlinear octupole and
hexadecapole TIs, hence our findings suggest new directions for exploring
solitons in a broad range of TIs.

\section{Results}

\subsection{The tight-binding model and circuit lattice}

\begin{figure}[t]
\centering
\includegraphics[width=8.6cm]{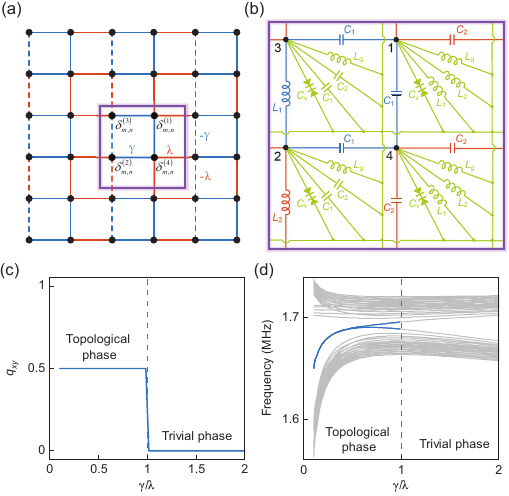}
\caption{\textbf{Schematic representation of the nonlinear quadrupole
topological insulator (NLQTI) and its topological properties in the linear limit.}
(a) The tight-binding lattice model of the NLQTI, where $\protect\gamma $
and $\protect\lambda $ represent the intracell and intercell hopping
strengths, respectively. Solid and dashed lines designate positive and
negative hoppings, respectively. Parameters $\protect\delta %
_{m,n}^{(1,2,3,4)}$ denote the amplitude-dependent onsite energies in the $%
(m,n)$-th unit cell. (b) The unit cell of the electric-circuit lattice
realizing the NLQTI. Blue and red circuit elements realize the intracell and
intercell hoppings, respectively, while yellow-green grounded circuit
elements determine the onsite energies. (c) The dependence of the quadrupole
moment $q_{x,y}$ on the dimerization ratio $\protect\gamma /\protect\lambda $%
. (d) The frequency spectrum of the lattice with open boundary conditions in
the $x$ and $y$ directions, plotted as a function of $\protect\gamma /%
\protect\lambda $. The blue curve indicates the corner states, while the
gray curves denote the bulk and edge ones. The red lines in (c) and (d) mark
the phase-transition boundary.}
\label{fig1}
\end{figure}

The tight-binding lattice supporting NLQTIs is presented in Fig. \ref{fig1}%
(a). Each unit cell consists of four lattice sites, with the $\left(m,n\right)$-th unit
cell indicated by the purple square. The intracell and intercell hopping
strengths are denoted as $\gamma $ and $\lambda $, respectively, with solid
and dashed lines designating positive and negative hoppings. Unlike the
linear QTI \cite{science357-61}, in Fig. \ref{fig1}(a) it is shown that the
onsite energies in our NLQTI, $\delta _{m,n}^{(j)}$, $j=1,2,3,4$, depend on
the wave functions $\psi _{m,n}^{(j)}$ at the respective sites. In photonic
and bosonic-gas systems, these dependences are typically represented by
quadratic forms \cite{RMP83-247,PR463-1,RPP75-086401,QF4-9} (see further
details in \ref{app_bosonic_gas} for the model of the interacting
bosonic gas). Here we realize this tight-binding model, using the
electric-circuit lattice shown in Fig.~\ref{fig1}(b). The circuit elements $%
\left( C_{1},L_{1}\right) $ and $\left( C_{2},L_{2}\right) $ determine the
intracell and intercell hoppings, respectively. To implement the
amplitude-dependent onsite energies, we employ two types of grounded
nonlinear oscillators. Circuit nodes $1$ and $4$ are connected to inductors $%
L_{\text{g}}$, $L_{1}$, $L_{2}$ and common-cathode diodes $C_{\text{v}}$,
while nodes $2$ and $3$ are connected to capacitors $C_{1}$ and $C_{2}$,
along with inductors $L_{\text{g}}$ and common-cathode diodes $C_{\text{v}}$%
. The common-cathode diodes function as voltage-dependent capacitors, with
capacitance
\begin{equation}
C\left( v\right) =\frac{C_{\text{L}}}{\left( 1+\left\vert v/v_{0}\right\vert \right) ^{M}},  \label{Cv}
\end{equation}
where $v$ is the voltage amplitude, with constants chosen here as $C_{\text{L%
}}=8.6~\text{nF}$, $v_{0}=1.7$, and $M=0.3$ (see details in \ref%
{app_modeling}).

The voltages at the circuit nodes in the $(m,n)$-th unit cell, denoted as $%
\psi _{m,n}^{(j)}$ with $j=1,2,3,4$, satisfy the following equation (for the
detailed derivation, see \ref{app_derivation}):
\begin{equation}
\mathrm{i}\frac{d\mathbf{\Psi }}{dt}=\mathbf{H\Psi },  \label{eq}
\end{equation}%
where $\mathbf{\Psi }=\left[ \psi _{m,n}^{(1)},\psi _{m,n}^{(2)},\psi
_{m,n}^{(3)},\psi _{m,n}^{(4)}\right] ^{\mathcal{T}}$, with $\mathcal{T}$
denoting the transpose of the vector. The Hamiltonian $\mathbf{H}$ is
defined as
\begin{equation}
\mathbf{H}=\begin{pmatrix} \delta_{m,n}^{(1)} & 0 & \mathbf{C}_{1,0} &
\mathbf{C}_{0,1} \\ 0 & \delta_{m,n}^{(2)} & -\mathbf{C}_{0,-1} &
\mathbf{C}_{-1,0} \\ \mathbf{C}_{-1,0} & -\mathbf{C}_{0,1} &
\delta_{m,n}^{(3)} & 0 \\ \mathbf{C}_{0,-1} & \mathbf{C}_{1,0} & 0 &
\delta_{m,n}^{(4)} \end{pmatrix},  \label{H}
\end{equation}%
where $\mathbf{C}_{k,l}=\gamma +\lambda \mathbf{T}_{k,l}$ and the discrete
translational operator is defined as $\mathbf{T}_{k,l}\psi _{m,n}^{\left(
j\right) }=\psi _{m+k,n+l}^{\left( j\right) }$. In Eq. (\ref{H}), the
above-mentioned amplitude-dependent onsite energies are $\delta
_{m,n}^{(j)}=\delta _{0}+g\left( \psi _{m,n}^{(j)}\right) $, where $\delta
_{0}=\left[ 1-\frac{C_{1}+C_{2}}{C_{\text{L}}}+\frac{L_{\text{g}}}{2}\left(
\frac{1}{L_{1}}+\frac{1}{L_{2}}\right) \right] \omega _{0}$ and $g\left(
\psi _{m,n}^{(j)}\right) =-\frac{C_{\text{v}}\left( \psi _{m,n}^{(j)}\right)
-C_{\text{L}}}{2C_{\text{L}}}\omega _{0}$, with $\omega _{0}=1/\sqrt{L_{%
\text{g}}C_{\text{L}}}$ and $C_{\text{v}}\left( \psi _{m,n}^{(j)}\right) $
defined as per Eq. (\ref{Cv}), with $v$ substituted by the respective $%
\psi _{m,n}^{(j)}$. The intracell and intercell hopping
strengths are $\gamma =\frac{C_{1}}{2C_{\text{L}}}\omega _{0}=\frac{L_{\text{%
g}}}{2L_{1}}\omega _{0}$ and $\lambda =\frac{C_{2}}{2C_{\text{L}}}\omega
_{0}=\frac{L_{\text{g}}}{2L_{2}}\omega _{0}$, respectively. We set $L_{\text{%
g}}=1~\mathrm{\mu H}$ (recall $\mathrm{H}\ $is the inductance unit, Henry)
and adjust the values of $L_{1}$ and $L_{2}$ accordingly, while determining
the values of the other circuit elements. In the linear limit, with $g=0$, this circuit configuration
satisfies two anti-commuting reflection symmetries, $m_{x}$ and $m_{y}$,
which quantize the quadrupole moment $q_{xy}$ to either $0$ or $\frac{1}{2}$
(for a detailed discussion, see \ref{app_linear_QI}). Specifically,
the system exhibits a topological quadrupole phase with $q_{xy}=\frac{1}{2}$
when $\gamma <\lambda $, and it is topologically trivial when $\gamma
>\lambda $, as illustrated in Fig. \ref{fig1}(c). The corner states emerge
due to the presence of the nontrivial bulk quadrupole moment, as highlighted
by the blue curves in Fig. \ref{fig1}(d), which displays the frequency
spectrum for a lattice with open boundary conditions in the $x$ and $y$
directions. The gray curves in Fig. \ref{fig1}(d) represent the bulk and
edge states.

\subsection{Nonlinear corner states and corner solitons: the theory and
experiment}

\begin{figure}[t]
\centering
\includegraphics[width=8.6cm]{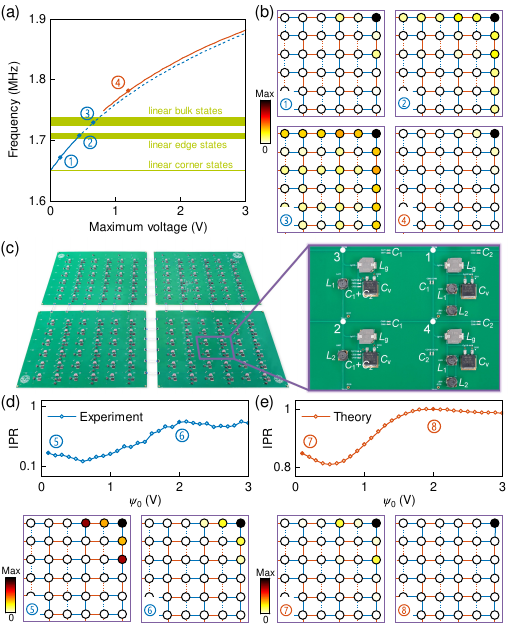}
\caption{\textbf{Nonlinear topological corner states and topologically
trivial corner solitons in the NLQTI.} (a) The dependence of the
eigenfrequency on the maximum voltage in the lattice. The solid and dashed
blue curves represent localized and delocalized nonlinear corner states,
respectively, while the red curve pertains to the corner solitons.
The frequencies corresponding to the linear corner, edge, and bulk states are indicated by the green regions.
(b) Absolute values of the voltage distributions for the states labeled in (a),
normalized to their respective maximum values. (c) The side view of the
fabricated PCB, with the inset exhibiting the unit cell of the circuit
lattice. The circuit elements are labeled in the inset. (d,e) Experimental
and theoretical IPRs of the voltage distributions (see Eq. (\protect\ref{IPR}%
)) at $t=9~\mathrm{\protect\mu s}$ for different initial voltages. Plots
marked as \textcircled{5} through \textcircled{8} display the voltage
distributions for the respective values of $\protect\psi _{0}$ in panels (d)
and (e).}
\label{fig3}
\end{figure}

We set $L_{1}=150\,~\mathrm{\mu }\text{H}$ and $L_{2}=15~\mathrm{\mu }$H,
defining $\mathbf{\Psi }=\mathbf{\Phi }e^{-\mathrm{i}2\pi ft}$. The
eigenfrequencies $f$ and eigenstates $\mathbf{\Phi }$ of the nonlinear
quadrupole circuit, governed by Eqs. (\ref{eq})-(\ref{H}), were calculated
using Newton's algorithm (see \ref{app_localization}).
Since the four linear corner states are equivalent to one another, in terms of both eigenfrequencies and 
eigenstates (see Fig. \ref{fig_linear}), we focus on the nonlinear corner states and corner solitons localized 
at the upper-right corner of the lattice, as a relevant example.

Figure \ref{fig3}(a) displays the relationship between the eigenfrequency and maximum voltage.
For comparison, the frequencies corresponding to the linear corner, edge, and bulk states are indicated by the green regions.
The nonlinear corner states (plotted by the blue curve) bifurcate
from the linear topological corner mode and remain well localized (see state \textcircled{1} in Fig. \ref{fig3}(b)).
A similar phenomenon is observed in arrays of evanescently coupled waveguides with all-positive hoppings \cite{nphys17-995}.
As the corner states are sublattice-polarized ones \cite%
{SP18-208}, the onsite energies can be approximated as $\delta
_{m,n}^{(1)}=\delta _{0}+g\left( \phi _{2,2}^{(1)}\right) $ and $\delta
_{m,n}^{(2,3,4)}=\delta _{0}$, where $\phi _{2,2}^{(1)}$ is the component of
$\mathbf{\Phi }$ representing the voltage at site $1$ of the $(2,2)$-th unit
cell, i.e., at the upper-right lattice corner. The reflection symmetries $m_{x}$ and $%
m_{y}$ are broken, resulting in the fact that the Wannier-state centers no
longer satisfy the relations inherent from linear QTIs---specifically, one
now has $v_{x,y}^{-}\neq -v_{x,y}^{+}\mod 1$. Although the Wannier-sector
polarizations $p_{y}^{v_{x}^{\pm }}$ and $p_{x}^{v_{y}^{\pm }}$ are no
longer quantized, the quadrupole moment $q_{xy}$ remains strictly positive,
unlike its nearly zero value for a trivial insulator (see \ref%
{app_topo_property}) \cite{LSA10-164,nphys17-995}. Thus, the nonvanishing
quadrupole invariant indicates that the corner states are topologically
nontrivial modes, persisting in the presences of the onsite nonlinearity.

As the maximum voltage increases and the frequency enters the band of linear
edge states, the nonlinear corner states hybridize with the edge ones and
become delocalized (see Fig. \ref{fig3}(a) and state \textcircled{2} in Fig. \ref{fig3}(b)).
The further increase in the eigenfrequency of the nonlinear corner states leads to additional delocalization through the hybridization
with bulk modes (see Fig. \ref{fig3}(a) and state \textcircled{3} in Fig. \ref{fig3}(b)). When the frequency continues to increase and enters
the semi-infinite gap, the nonlinear corner states remain delocalized.
This behavior is consistent with the previous observations in waveguide arrays \cite{nphys17-995}.
Considering the delocalized field distributions, the previous approximation for
the onsite energy becomes invalid, and the periodicity of the lattice is broken.
As a result, the original
definition of the quadrupole moment $q_{xy}$ is no longer applicable. Due to
the competition between the onsite self-interaction and intersite hoppings,
localized states do not exist in this moderately nonlinear regime.

In the regime of strong nonlinearity, a new branch corresponding to corner
solitons emerges, indicated by the red curve in Fig. \ref{fig3}(a).
These solitons arise at a nonzero voltage because they become too weakly localized at low voltages (see Appendix D.1), 
therefore the numerical method is unable to capture poorly localized solutions for a finite lattice, 
similar to the results reported in Ref. \cite{PRB102-115411}.
As these corner solitons are generally strongly localized at the corner sites and lack the
sublattice polarization (see state \textcircled{4} in Fig. \ref{fig3}(b)), the previous approximation, $%
\delta ^{(1)}=\delta _{0}+g\left(\phi _{2,2}^{(1)}\right)$ and $\delta
^{(2,3,4)}=\delta _{0}$, is no longer valid. Instead, when the onsite
energies dominate over the differences between the intracell and intercell
hopping strengths, the NLQTI may be approximated as a nonlinear lattice with
equal intracell and intercell hopping strengths $\frac{\gamma +\lambda }{2}$
(see \ref{app_topo_property}). This lattice exhibits no
dimerization and supports conventional self-trapped states residing at the
lattice edges, commonly referred to as surface solitons, which are
topologically trivial modes \cite{RMP83-247,PR463-1,RPP75-086401}. Here, as
such states are located at the corner, which is the junction of two edges,
we call this special type of the surface modes \textit{corner solitons} \cite%
{LSA10-164,nphys17-995}. As the maximum voltage increases to a sufficiently
high level, Eqs. (\ref{eq})-(\ref{H}) reduce to $\mathrm{i}\frac{d\psi
_{2,2}^{(1)}}{dt}=\delta _{2,2}^{(1)}\left( \psi _{2,2}^{(1)}\right) \psi
_{2,2}^{(1)}$, governing the dynamics of the nonlinear single lattice site.

As both the localized nonlinear corner states and topologically trivial
corner solitons are dynamically stable (see Fig. \ref%
{fig_stability_corner_state} and \ref{app_corner_stability}), we
have verified their existence experimentally, through quench dynamics (see
details in \ref{app_exp}) \cite{CP8-342}. Figure \ref{fig3}(c)
shows the circuit lattice fabricated as a printed circuit board (PCB), which
carries a $6\times 6$ array of unit cells. The inset displays the unit
cell of the circuit lattice, with circuit elements labeled in it.
For the experiment, the initial state at $t=0$ was prepared as
$\psi_{m,n}^{(i)}(t=0) = \psi_{0} \delta_{m,2} \delta_{n,2} \delta_{i,1},$
with a nonzero voltage applied solely at the upper-right corner site. The voltage distribution at $t \ge 0$ was recorded 
during the temporal evolution, using an oscilloscope. Subsequently, the envelope of the temporal voltage signals corresponding to $\psi_{m,n}^{(i)}(t)$ was extracted.

In the weakly nonlinear regime, when $\psi _{0}$ is
small, the initial state can be expanded into a set of the eigenstates of
the respective linear model: $\mathbf{\Psi }\left(t=0\right)=\sum_{n}c_{n}\mathbf{\Phi }%
_{n}$, where $c_{n}$ are expansion coefficients. Thus we found $\mathbf{\Psi
}\left(t\right)=\sum_{n}c_{n}e^{-\mathrm{i}2\pi f_{n}t}\mathbf{\Phi }_{n}$, with $f_{n}$
being the eigenfrequency of the $n$-th eigenstate \cite{PRB104-235420}. As
the initial state largely overlaps with the corner one, the expansion
coefficient corresponding to the corner state is close to $1$, resulting in
a localized voltage distribution with sublattice polarization as a result of the temporal evolution. Conversely,
when $\psi _{0}$ attains a medium value, corresponding to the moderately
nonlinear regime, we expect a delocalized voltage distribution due to the
absence of localized eigenstates. When $\psi _{0}$ becomes sufficiently
large (in the strongly nonlinear regime), the initial state predominantly
overlaps with the eigenstate of the corner soliton. Particularly, in the
ultra-strongly nonlinear regime, the initial state perfectly matches the
single-site corner soliton, leading to the sharply localized topologically
trivial voltage distributions.

We define the inverse participation ratio (IPR) as
\begin{equation}
\text{IPR}=\frac{\sum\limits_{m,n,i}\left\vert \psi _{m,n}^{(i)}\right\vert
^{4}}{\left( \sum\limits_{m,n,i}\left\vert \psi _{m,n}^{(i)}\right\vert
^{2}\right) ^{2}},  \label{IPR}
\end{equation}%
which measures the localization degree of the field distribution, with a
higher IPR indicating stronger localization. Theoretically, the IPR should be
evaluated after a sufficiently long evolution time. However, due to inevitable circuit
dissipations in experiments, we extract the voltage distributions at $t=9~\mathrm{\mu s}$,
with the corresponding IPR results for the experimental voltage distributions
shown in Fig. \ref{fig3}(d). The insets \textcircled{5} and %
\textcircled{6} present the voltage distributions for two representative values of $\psi _{0}$.
The theoretical results, derived from the numerical
solutions of Eqs. (\ref{eq})-(\ref{H}), are displayed in Fig. \ref{fig3}(e).
For comparison, the measurement result and theoretical predication of voltage distribution in
the moderately nonlinear regime are shown in Fig. \ref{fig_vol_medium}(a).

From the results, we observe that the voltage distributions are localized for both small and large initial voltages, while they appear comparatively delocalized at medium initial voltage values. This localization-delocalization-localization transition indicates that nonlinear corner states and corner solitons are excited in the weakly and strongly nonlinear regimes, respectively, whereas no localized states exist in the medium-nonlinearity regime (see Fig. \ref{fig_illustration} for an illustration of the transition and \ref{summary} for a relevant discussion).
Although the circuit dissipation causes deviations in the specific values of the experimental result, it primarily induces an overall reduction in the IPRs, which are measured experimentally, and does not alter the overall trend, which agrees well with the theoretical prediction. This implies that the selection of the time moment $t = 9~\mathrm{\mu s}$ is appropriate for observing the localization-delocalization-localization transition and for validating the existence of nonlinear corner states and corner solitons. Our conclusion is further supported by the result obtained for longer evolution times, where the localization-delocalization-localization transition is again observed (see Fig. \ref{fig_long} and \ref{app_quench}).

\subsection{Bulk solitons}

\begin{figure}[t]
\centering
\includegraphics[width=8.6cm]{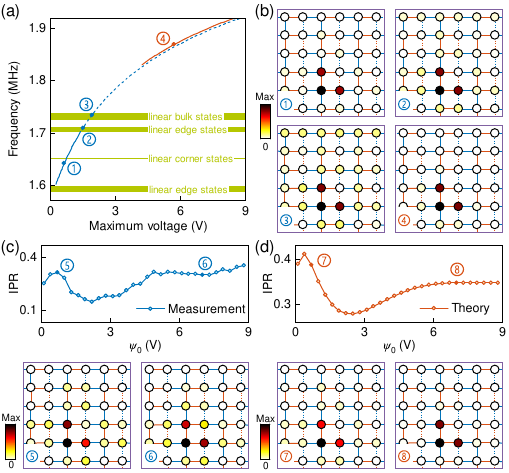}
\caption{\textbf{Bulk solitons in the NLQTI.} (a) The eigenfrequency versus the
maximum voltage in the lattice.
The solid blue and red curves represent two distinct types of bulk solitons, both of which are localized states.
The dashed blue curve indicates the delocalized states that arise from one type of the bulk soliton.
For comparison, the frequencies corresponding to the linear corner, edge, and bulk states are indicated by the green regions.
(b) Absolute values of the normalized voltage distributions for the quasi-antisymmetric
triangular solitons labeled in (a). (c,d) Experimentally measured and
theoretically predicted IPRs of the voltage distributions at $t=8~\mathrm{%
\protect\mu }\text{s}$ for different initial voltages, with insets
\textcircled{5} -- \textcircled{8} showing the voltage distributions at
representative values of $\protect\psi _{0}$.}
\label{fig4}
\end{figure}

We then focus on the solitons residing in the bulk of the NLQTI. In the
anti-continuum limit \cite{Aubry}, we identify solution branches for the bulk
solitons, beginning with the ultimate case of the lattice without intersite
hopping, wherein only two states are expected to be observed experimentally (see \ref%
{app_bulk} for a detailed discussion).

Figure \ref{fig4}(a) displays the frequency spectra for these two types of bulk solitons.
The frequencies corresponding to the linear corner, edge, and bulk states are indicated by the green regions.
Although the bulk solitons are expected to approach the bulk bands in the linear limit \cite{arxiv},
the termination of the soliton branches is, once again, explained by the weak localization of the solitons at low voltages (see \ref{app_bulk_existence})
and the limitation of the numerical algorithm.
In the middle finite bandgap between two
bands of the linear edge states, a single type exists, represented by the
solid blue curve in Fig. \ref{fig4}(a).
This soliton features a triangular
quasi-antisymmetric profile, characterized by the relation $\psi _{0,0}^{\left( 1\right) }\approx -\psi
_{1,0}^{\left( 3\right) }=-\psi _{0,1}^{\left( 4\right) }$ (see state \textcircled{1} in Fig. \ref%
{fig4}(b)).
Once the frequency of this soliton enters the edge band, it
becomes a delocalized state, as indicated by the dashed blue curve in Fig. \ref{fig4}(a) and state \textcircled{2} in Fig. \ref%
{fig4}(b). A further increase in frequency results in hybridization with bulk states, causing the bulk soliton to become more delocalized, as shown by state \textcircled{3} in Fig. \ref{fig4}(b).
Additionally, a new branch for another type of the bulk
solitons emerges in the semi-infinite gap, as shown by the solid red curve in
Fig. \ref{fig4}(a). These solitons also exhibit quasi-antisymmetric profiles
and remain localized (see state \textcircled{4} in Fig. \ref{fig4}(b)).
They correspond to conventional lattice solitons
residing in the bulk of the lattice, getting more localized as the
nonlinearity strengthens \cite{RMP83-247,PR463-1,RPP75-086401}.
Moreover, it is relevant to emphasize that both types of bulk solitons can still be found even if one swaps the intracell 
and intercell hopping strengths, thereby transforming the NLQTI into a lattice that is a topologically trivial QTI in the linear limit.

We explored the quench dynamics to experimentally verify the existence of the two theoretically predicted types of bulk solitons. Unlike the single-site excitation shown in Fig. \ref{fig3}, we prepared an initial state at $t = 0$ in a triangular quasi-antisymmetric form, with $\psi_{0,0}^{(1)} = -\psi_{1,0}^{(3)} = -\psi_{0,1}^{(4)} \equiv \psi_{0}$ and zero voltages at all other sites. The voltage distributions resulting from the temporal evolution were recorded, and voltage envelopes corresponding to $\psi_{m,n}^{(i)}(t)$ were extracted. At the selected moment $t = 8~\mathrm{\mu s}$, the results shown in Figs. \ref{fig4}(c)-(d) reveal that the voltage distributions are localized for both small and large initial voltages, while they appear comparatively delocalized at medium initial voltage values (see Fig. \ref{fig_vol_medium}(b) for voltage distributions in the moderately nonlinear regime). 
The slight decrease in the IPRs near $\psi_{0} = 0$ is attributed to a mismatch between the initial excitation and the established soliton shape (see further details in \ref{app_quench}).
Additionally, the overall reduction in the experimentally measured IPRs is due to the circuit dissipation. Nonetheless, the consistent localization-delocalization-localization transition, which is revealed by the experimental measurements and theoretical prediction alike, confirms that one type of the bulk solitons occupies the middle finite bandgap under the action of weak nonlinearity, while the other type emerges in the semi-infinite gap under strong nonlinearity. As expected, no localized solitons are found in the system with medium nonlinearity strength. Figure \ref{fig_illustration} illustrates this transition, and \ref{summary} provides a relevant discussion.

\section{Discussion}

It is relevant to summarize unique phenomena exhibited by the NLQTI proposed
and implemented in the present work. First, in the nonlinear Wannier-type
HOTI proposed in Ref. \cite{PRL123-053902}, localized states arise solely as
nonlinearity-induced topological corner states. In contrast, our NLQTI
demonstrates the nonlinearity-controlled
localization-delocalization-localization transition, allowing the setup to
switch between the nonlinear topological corner states and topologically
trivial corner solitons. Second, in the nonlinear Wannier-type HOTIs, corner
states are inherently coupled to the edge modes \cite{LSA10-164} or to bulk
states when the lattice exhibits weak dimerization \cite{nphys17-995}.
Conversely, the corner states in QTIs are sharply separated from other
states, admitting the existence of the well-localized nonlinear corner
states and bulk solitons, even when the lattice dimerization is weak
(see further details in \ref{app_comparison}).

Thus, we have introduced the concept of NLQTIs and demonstrated their
experimental realization in an electric-circuit lattice. Through\ the
quench dynamics, we have observed the nonlinear topological corner states
and topologically trivial corner solitons in the weakly and strongly
nonlinear regimes, respectively. Further, this work reveals the formation of
two distinct types of the bulk solitons: one type exists in the middle
finite bandgap under the action of weak nonlinearity, while the other one
populates the semi-infinite gap under strong nonlinearity. Eventually, the
present work reports the advancement in the studies of nonlinear HOTIs,
establishing another member of the nonlinear HOTI family and suggesting
new avenues for exploring novel solitons across a broad range of TIs.

\section*{CRediT authorship contribution statement}

\textbf{Rujiang Li}: Conceptualization, Data curation, Formal analysis, Funding acquisition, Investigation,
Methodology, Project administration, Resources, Software, Supervision, Validation, Visualization,
Writing -- original draft, Writing -- review \& editing.
\textbf{Wencai Wang}: Investigation, Visualization.
\textbf{Yongtao Jia}: Resources.
\textbf{Ying Liu}: Funding acquisition, Project administration, Resources.
\textbf{Pengfei Li}: Formal analysis, Methodology.
\textbf{Boris~A.~Malomed}: Formal analysis, Supervision, Writing -- review \& editing.

\section*{Declaration of competing interest}

The authors declare that they have no known competing financial
interests or personal relationships that could have appeared to
influence the work reported in this paper.

\section*{Data availability}

Data will be made available on request.

\section*{Acknowledgements}

The authors would like to thank Haoran Xue and Jian-Hua Jiang for fruitful
discussions.
R.L. and W.W. were sponsored by the National Key Research and Development
Program of China (Grant No. 2022YFA1404902), National Natural Science
Foundation of China (Grant No. 12104353), and Fundamental Research Funds for
the Central Universities (Grant No. QTZX25086). Y.L. was sponsored by the
National Natural Science Foundation of China (NSFC) under Grant No. 62271366
and the 111 Project. P.L. was sponsored by the National Natural Science
Foundation of China (11805141) and Basic Research Program of Shanxi Provence
(202303021211185). The work of B.A.M. was supported, in part, by the Israel
Science Foundation through grant No. 1695/22.
Numerical calculations performed in this work were supported by
the High-Performance Computing Platform of Xidian University.

\appendix

\section{The model of the interacting bosonic gas\label{app_bosonic_gas}}

In this section, we construct the model for the interacting boson gas and
demonstrate that it effectively emulates the NLQTI. We consider a gas of
identical bosons hopping on a two-dimensional (2D) lattice. The respective
Hamiltonian,
\begin{equation}
\hat{H}=\hat{H}_{\text{kin}}+\hat{H}_{\text{int}},  \label{app_H}
\end{equation}%
includes the kinetic term,
\begin{eqnarray}
\hat{H}_{\text{kin}} &=&\sum_{m,n}\left[ \gamma \left( c_{m,n}^{(1)\dagger
}c_{m,n}^{(3)}+c_{m,n}^{(2)\dagger }c_{m,n}^{(4)}+\text{H.c.}\right) \right.
\nonumber \\
&&\left. +\gamma \left( c_{m,n}^{(1)\dagger
}c_{m,n}^{(4)}-c_{m,n}^{(2)\dagger }c_{m,n}^{(3)}+\text{H.c.}\right) \right.
\nonumber \\
&&\left. +\lambda \left( c_{m,n}^{(1)\dagger
}c_{m+1,n}^{(3)}+c_{m,n}^{(4)\dagger }c_{m+1,n}^{(2)}+\text{H.c.}\right)
\right.  \nonumber \\
&&\left. +\lambda \left( c_{m,n}^{(1)\dagger
}c_{m,n+1}^{(4)}-c_{m,n}^{(3)\dagger }c_{m,n+1}^{(2)}+\text{H.c.}\right)
\right.  \nonumber \\
&&\left. +\sum_{j}\delta _{i}c_{m,n}^{(j)\dagger }c_{m,n}^{(j)}\right] ,
\label{eq_S2}
\end{eqnarray}%
where H.c. stands for the Hermite conjugate expression, $c_{m,n}^{(j)\dagger
}$ and $c_{m,n}^{(j)}$ are the bosonic creation and annihilation operators,
respectively, at site $j$ in the unit cell $(m,n)$, with $j=1,2,3,4$, which
obey the canonical commutation relations:
\begin{equation}
\left[ c_{m,n}^{(j)},c_{m^{\prime },n^{\prime }}^{(j^{\prime } )\dagger}%
\right] =\delta _{m,m^{\prime }}\delta _{n,n^{\prime }}\delta _{j,j^{\prime
}}.
\end{equation}%
Here, $\gamma $ is the hopping amplitude for particles between adjacent
sites in the same unit cell, while $\lambda $ is the amplitude of hopping
between nearest-neighbor unit cells. The negative signs in Eq. (\ref{eq_S2})
correspond to the gauge choice which implies the $\pi $-flux threading
through a plaquette (including the unit cell itself). Further, $\delta _{j}$
with $j=1,2,3,4$ represent the energy offsets between sites within one unit
cell. The interaction Hamiltonian in Eq. (\ref{H}) represents the onsite
interactions between particles:
\begin{equation}
\hat{H}_{\text{int}}=\frac{g}{2}\sum_{m,n,j}\hat{n}_{m,n}^{(j)}\left( \hat{n}%
_{m,n}^{(j)}-1\right) ,
\end{equation}%
where $\hat{n}_{m,n}^{(j)}$ is the number operator counting the bosons at
site $j$ in unit cell $(m,n)$, and $g$ is the strength of the onsite
interaction between the particles, while we neglect the interaction between
particles on different sites. Using the commutation relations, the
interaction Hamiltonian can be rewritten as:
\begin{equation}
\hat{H}_{\text{int}}=\frac{g}{2}\sum_{m,n,j}c_{m,n}^{(j)\dagger
}c_{m,n}^{(j)\dagger }c_{m,n}^{(j)}c_{m,n}^{(j)}.
\end{equation}

Solving the many-body interacting problem for Hamiltonian $\hat{H}$ from Eq.
(\ref{H}) is a challenging task. In the mean-field approximation, the
macroscopic state, i.e., $\left\vert \Phi \right\rangle $, is chosen as the
tensor product of Glauber coherent states $\left\vert \Phi \right\rangle
=\otimes _{m,n,j}\left\vert \psi _{m,n}^{\left( j\right) }\right\rangle
_{\left( \text{GCS}\right) }$, where $\left\vert \psi _{m,n}^{\left(
j\right) }\right\rangle _{\left( \text{GCS}\right) }$ describes only a
single site and it is an eigenstate of the annihilation operator with $%
c_{m,n}^{\left( j\right) }\left\vert \psi _{m,n}^{\left( j\right)
}\right\rangle _{\left( \text{GCS}\right) }=\psi _{m,n}^{\left( j\right)
}\left\vert \psi _{m,n}^{\left( j\right) }\right\rangle _{\left( \text{GCS}%
\right) }$ \cite{PR129-959}. With $H=\left\langle \Phi \left\vert \hat{H}%
\right\vert \Phi \right\rangle $, the respective semi-classical Hamiltonian
can be written as:
\begin{eqnarray}
H &=&\sum_{m,n}\left[ \gamma \left( \psi _{m,n}^{(1)\ast }\psi
_{m,n}^{(3)}+\psi _{m,n}^{(2)\ast }\psi _{m,n}^{(4)}+\psi _{m,n}^{(3)\ast
}\psi _{m,n}^{(1)}+\psi _{m,n}^{(4)\ast }\psi _{m,n}^{(2)}\right) \right.
\nonumber \\
&&+\gamma \left( \psi _{m,n}^{(1)\ast }\psi _{m,n}^{(4)}-\psi
_{m,n}^{(2)\ast }\psi _{m,n}^{(3)}+\psi _{m,n}^{(4)\ast }\psi
_{m,n}^{(1)}-\psi _{m,n}^{(3)\ast }\psi _{m,n}^{(2)}\right)  \nonumber \\
&&+\lambda \left( \psi _{m,n}^{(1)\ast }\psi _{m+1,n}^{(3)}+\psi
_{m,n}^{(4)\ast }\psi _{m+1,n}^{(2)}+\psi _{m+1,n}^{(3)\ast }\psi
_{m,n}^{(1)}+\psi _{m+1,n}^{(2)\ast }\psi _{m,n}^{(4)}\right)  \nonumber \\
&&+\lambda \left( \psi _{m,n}^{(1)\ast }\psi _{m,n+1}^{(4)}-\psi
_{m,n}^{(3)\ast }\psi _{m,n+1}^{(2)}+\psi _{m,n+1}^{(4)\ast }\psi
_{m,n}^{(1)}-\psi _{m,n+1}^{(2)\ast }\psi _{m,n}^{(3)}\right)  \nonumber \\
&&+\sum_{j}\delta _{j}\left\vert \psi _{m,n}^{(j)}\right\vert ^{2} \nonumber \\
&& \left. +\frac{g}{2}\left( \left\vert \psi _{m,n}^{(1)}\right\vert ^{4}+\left\vert
\psi _{m,n}^{(2)}\right\vert ^{4}+\left\vert \psi _{m,n}^{(3)}\right\vert
^{4}+\left\vert \psi _{m,n}^{(4)}\right\vert ^{4}\right) \right] .
\label{H2}
\end{eqnarray}%
By rearranging the terms in the above expression, this Hamiltonian can be
rewritten as:
\begin{eqnarray}
H &=&\sum_{m,n}\left[\left( \gamma \psi _{m,n}^{(3)}+\gamma \psi
_{m,n}^{(4)}+\lambda \psi _{m+1,n}^{(3)}+\lambda \psi _{m,n+1}^{(4)}\right)
\psi _{m,n}^{(1)\ast } \right. \nonumber \\
&&+\left( -\gamma \psi _{m,n}^{(3)}+\gamma \psi _{m,n}^{(4)}+\lambda \psi
_{m-1,n}^{(4)}-\lambda \psi _{m,n-1}^{(3)}\right) \psi _{m,n}^{(2)\ast }
\nonumber \\
&&+\left( \gamma \psi _{m,n}^{(1)}-\gamma \psi _{m,n}^{(2)}+\lambda \psi
_{m-1,n}^{(1)}-\lambda \psi _{m,n+1}^{(2)}\right) \psi _{m,n}^{(3)\ast }
\nonumber \\
&&+\left( \gamma \psi _{m,n}^{(1)}+\gamma \psi _{m,n}^{(2)}+\lambda \psi
_{m+1,n}^{(2)}+\lambda \psi _{m,n-1}^{(1)}\right) \psi _{m,n}^{(4)\ast }
\nonumber \\
&&+\sum_{j}\delta _{j}\left\vert \psi _{m,n}^{(j)}\right\vert ^{2} \nonumber \\
&&\left.+\frac{g}{2}\left( \left\vert \psi _{m,n}^{(1)}\right\vert ^{4}+\left\vert
\psi _{m,n}^{(2)}\right\vert ^{4}+\left\vert \psi _{m,n}^{(3)}\right\vert
^{4}+\left\vert \psi _{m,n}^{(4)}\right\vert ^{4}\right) \right] .
\label{H3}
\end{eqnarray}%
The respective equations of motion,
\begin{equation}
\mathrm{i}\frac{d\psi _{m,n}^{(j)}}{dt}=\frac{\delta H}{\delta \psi
_{m,n}^{(j)\ast }},\quad j=1,2,3,4,
\end{equation}%
derived from Hamiltonian (\ref{H3}), take the following form:
\begin{eqnarray}
\mathrm{i}\frac{d\psi _{m,n}^{(1)}}{dt} &=&\delta _{1}\psi
_{m,n}^{(1)}+\gamma \psi _{m,n}^{(3)}+\gamma \psi _{m,n}^{(4)}+\lambda \psi
_{m+1,n}^{(3)}+\lambda \psi _{m,n+1}^{(4)}  \nonumber \\
&&+g\left\vert \psi _{m,n}^{(1)}\right\vert ^{2}\psi _{m,n}^{(1)}, \\
\mathrm{i}\frac{d\psi _{m,n}^{(2)}}{dt} &=&\delta _{2}\psi
_{m,n}^{(2)}+\gamma \psi _{m,n}^{(4)}-\gamma \psi _{m,n}^{(3)}+\lambda \psi
_{m-1,n}^{(4)}-\lambda \psi _{m,n-1}^{(3)}  \nonumber \\
&&+g\left\vert \psi _{m,n}^{(2)}\right\vert ^{2}\psi _{m,n}^{(2)}, \\
\mathrm{i}\frac{d\psi _{m,n}^{(3)}}{dt} &=&\delta _{3}\psi
_{m,n}^{(3)}+\gamma \psi _{m,n}^{(1)}-\gamma \psi _{m,n}^{(2)}+\lambda \psi
_{m-1,n}^{(1)}-\lambda \psi _{m,n+1}^{(2)}  \nonumber \\
&&+g\left\vert \psi _{m,n}^{(3)}\right\vert ^{2}\psi _{m,n}^{(3)}, \\
\mathrm{i}\frac{d\psi _{m,n}^{(4)}}{dt} &=&\delta _{4}\psi
_{m,n}^{(4)}+\gamma \psi _{m,n}^{(2)}+\gamma \psi _{m,n}^{(1)}+\lambda \psi
_{m+1,n}^{(2)}+\lambda \psi _{m,n-1}^{(1)}  \nonumber \\
&&+g\left\vert \psi _{m,n}^{(4)}\right\vert ^{2}\psi _{m,n}^{(4)}.
\end{eqnarray}%
Setting $\delta _{1}=\delta _{2}=\delta _{3}=\delta _{4}$, these equations
describe the NLQTI, where the usual onsite energies, corresponding to the
cubic nonlinearity, are quadratic functions of the amplitudes at the
corresponding sites.

\section{The derivation of circuit equations for the nonlinear quadrupole
topological insulator\label{derivation}}

This section is organized as follows. In the first subsection, we introduce
a model with the common-cathode diodes. Then, in the second subsection, we
derive the circuit equations which take the form as the governing equations
of the NLQTI.

\subsection{The model with common-cathode diodes\label{app_modeling}}

\begin{figure}[t]
\centering
\includegraphics[width=4.5cm]{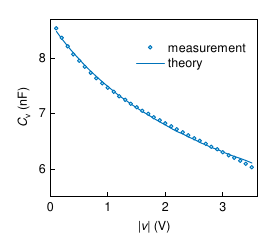}
\caption{\textbf{The capacitance-voltage relation of two parallel-connected
common-cathode diodes.} The curve represents theoretical results, while the
circular points correspond to data of the experimental measurements.}
\label{fig_varactor}
\end{figure}

We utilize two parallel-connected common-cathode diodes (V60DM45C) as the
onsite nonlinearity for the NLQTI. These diodes are mounted on the top and
bottom layers of the PCB, respectively. Similar to back-to-back varactor
diodes, the use of common-cathode diodes ensures identical responses in the
course of the two half-cycles of the AC voltage, eliminating the need for a
bias voltage \cite{nelectron1-178}. Although the capacitance of the
common-cathode diodes is time-dependent, our simulations and experimental
measurements show that the high-harmonic currents induced by the diodes are
negligible. Therefore, the capacitance of the common-cathode diodes can be
expressed as:
\begin{equation}
C=\mathrm{i}\frac{j_{\text{v}}}{\omega v},  \label{eq_C_reduce}
\end{equation}%
where $j_{\text{v}}$ and $v$ represent the amplitudes of the current and
voltage, respectively, and $\omega $ is the angular frequency of the applied voltage
signal. After obtaining the experimental data for the capacitance-voltage
relation of two parallel-connected common-cathode diodes, we fit the
capacitance-voltage curve to expression (\ref{Cv}) in the main text, i.e.,
\begin{equation}
C\left( v\right) =\frac{C_{\text{L}}}{\left( 1+\left\vert v/v_{0}\right\vert \right) ^{M}},
\label{C_model}
\end{equation}%
where $C_{\text{L}}$ is the capacitance at zero voltage, $v_{0}$ and $M$ are
constants, and $v$ represents the amplitude of the applied voltage. Figure %
\ref{fig_varactor} shows the capacitance-voltage relation of two
parallel-connected common-cathode diodes, with the curve representing the
theoretical results and the circular points corresponding to the
experimental data. The adopted parameters of the parallel-connected
common-cathode diodes are $C_{\text{L}}=8.6~\text{nF}$, $v_{0}=1.7$, and $%
M=0.3$. Thus, the common-cathode diodes function as variable capacitors,
with the capacitance depending on the voltage amplitude. We have also
performed experimental measurements of the common-cathode diodes at other
frequencies, concluding that Eq. (\ref{C_model}) remains valid across the
entire parameter range of our study. Extending this conclusion, when the
applied voltage is quasi-monochromatic with $V\left( t\right) =\frac{1}{2}%
v\left( t\right) e^{-\mathrm{i}\omega t}+\text{c.c.}$, the capacitance of
the common-cathode diodes can be written as
\begin{equation}
C\left( v\right) =\frac{C_{\text{L}}}{\left[ 1+\left\vert v\left( t\right) /v_{0} \right\vert%
\right] ^{M}},  \label{C_model_1}
\end{equation}%
where $v\left( t\right) $ is the slowly-varying envelope amplitude. In this
case, the capacitance of the common-cathode diodes varies with the voltage
envelope.

\subsection{The derivation of circuit equations\label{app_derivation}}

\begin{figure}[t]
\centering
\includegraphics[width=6.2cm]{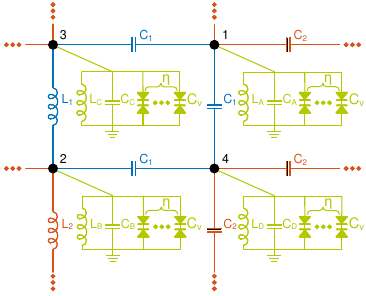}
\caption{\textbf{A schematic of the NQTI realized in the circuit lattice.}
For the clarity's sake, only the unit cell is shown.}
\label{fig_circuit}
\end{figure}

We aim to build a generic circuit lattice emulating the NLQTI. As shown in
Fig.~\ref{fig_circuit}, the system is built of nonlinear oscillators
composed of inductors $L_{\text{A},\text{B},\text{C},\text{D}}$, capacitors $%
C_{\text{A},\text{B},\text{C},\text{D}}$, and common-cathode diodes with
capacitance $C_{\text{v}}$. To enhance the nonlinear response, each
oscillator includes $J$ parallel-connected common-cathode diodes (i.e.,
the number of common-cathode diodes is $J$).
The oscillators are interconnected by coupling elements $C_{1}$,
$C_{2}$, $L_{1}$, and $L_{2}$. The common-cathode diode acts as a
voltage-dependent capacitor, with the total capacitance $C_{\text{v}}=C_{%
\text{L}}+C_{\text{NL}},$ where $C_{\text{L}}$ and $C_{\text{NL}}$ are the
linear and nonlinear terms, respectively. The linear part $C_{\text{L}}$ is
constant, while for the time-harmonic voltage $V(t)=\frac{1}{2}ve^{-\mathrm{i%
}\omega t}+\text{c.c.}$, the nonlinear component is $C_{\text{NL}}=-C_{\text{%
L}}+\frac{C_{\text{L}}}{\left(1+\left\vert v/v_{0}\right\vert \right)^{M}}$, with constant parameters $v_{0}$
and $M$ (cf. Eq. (\ref{C_model_1})). The grounding capacitance and
inductance satisfy relations
\begin{equation}
C_{\text{A},\text{B},\text{C},\text{D}}=C_{%
\text{g}}+\Delta C_{\text{A},\text{B},\text{C},\text{D}}
\end{equation}
and
\begin{equation}
\frac{1}{L_{\text{A},\text{B},\text{C},\text{D}}}=\frac{1}{L_{\text{g}}}+\frac{1}{\Delta
L_{\text{A},\text{B},\text{C},\text{D}}},
\end{equation}
respectively.

Usually, the differences between $C_{\text{A}}$, $C_{\text{B}}$, $C_{\text{C}%
}$, and $C_{\text{D}}$ are small, and the nonlinear component of the diode
capacitance, $C_{\text{NL}}$, is small too. Therefore, the total grounding
capacitance may be approximated as $C_{\text{g}}+J C_{\text{L}}$.
Similarly, the differences between $L_{\text{A}}$, $L_{\text{B}}$, $L_{\text{%
C}}$, and $L_{\text{D}}$ are also small. Given that $C_{\text{g}}+J C_{%
\text{L}}\gg C_{1,2}$ and $L_{\text{g}}\ll L_{1,2}$, the voltages at the
circuit nodes can be expressed as $V(t)=\frac{1}{2}v(t)e^{-\mathrm{i}\omega
_{0}t}+\text{c.c.}$, where $v(t)$ is the slowly-varying complex amplitude
and
\begin{equation}
\omega _{0}=\frac{1}{\sqrt{L_{\text{g}}(C_{\text{g}}+J C_{\text{L}})}}
\end{equation}
is the carrier frequency. Accordingly, the nonlinear component of
the diode capacitance is given by
\begin{equation}
C_{\text{NL}}=-C_{\text{L}}+\frac{C_{%
\text{L}}}{\left[ 1+\left\vert v\left(t\right)/v_{0}\right\vert \right] ^{M}}.
\end{equation}

For the four circuit nodes in the $(m,n)$-th unit cell, the following
relations can be obtained using the Kirchhoff's law:
\begin{eqnarray}
&&C_{1}\frac{d^{2}V_{m,n}^{\left(3\right)}}{dt^{2}}
+C_{2}\frac{d^{2}V_{m+1,n}^{\left(3\right)}}{dt^{2}}
+C_{1}\frac{d^{2}V_{m,n}^{\left(4\right)}}{dt^{2}}
+C_{2}\frac{d^{2}V_{m,n+1}^{\left(4\right)}}{dt^{2}} \nonumber \\
&&-\left[\Delta C_{\text{A}}+2C_{1}+2C_{2}+J C_{\text{NL}}\left(v_{m,n}^{\left(1\right)} \right) \right]%
\frac{d^{2}V_{m,n}^{\left(1\right)}}{dt^{2}}
-\frac{V_{m,n}^{\left(1\right)}}{\Delta L_{\text{A}}} \nonumber \\
&=&\frac{V_{m,n}^{\left(1\right)}}{L_{\text{g}}}
+\left( C_{\text{g}}+J C_{\text{L}}\right) \frac{d^{2}V_{m,n}^{\left(1\right)}}{dt^{2}},  \label{circuit_eq_1} \\
&&C_{2}\frac{d^{2}V_{m-1,n}^{\left(4\right)}}{dt^{2}}
+C_{1}\frac{d^{2}V_{m,n}^{\left(4\right)}}{dt^{2}}
+\frac{V_{m,n}^{\left(3\right)}}{L_{1}}
+\frac{V_{m,n-1}^{\left(3\right)}}{L_{2}}  \nonumber \\
&&-\left[\Delta C_{\text{B}}+C_{1}+C_{2}+J C_{\text{NL}} \left(v_{m,n}^{\left(2\right)}\right) \right]
\frac{d^{2}V_{m,n}^{\left(2\right)}}{dt^{2}} \nonumber \\
&&-\left(\frac{1}{\Delta L_{\text{B}}}+\frac{1}{L_{1}}+\frac{1}{L_{2}}\right)V_{m,n}^{\left(2\right)}  \nonumber \\
&=&\frac{V_{m,n}^{\left(2\right)}}{L_{\text{g}}}
+\left( C_{\text{g}}+J C_{\text{L}}\right) \frac{d^{2}V_{m,n}^{\left(2\right)}}{dt^{2}},  \label{circuit_eq_2}
\end{eqnarray}
\begin{eqnarray}
&&C_{2}\frac{d^{2}V_{m-1,n}^{\left(1\right)}}{dt^{2}}
+C_{1}\frac{d^{2}V_{m,n}^{\left(1\right)}}{dt^{2}}
+\frac{1}{L_{1}}V_{m,n}^{\left(2\right)}
+\frac{1}{L_{2}}V_{m,n+1}^{\left(2\right)} \nonumber \\
&&-\left[\Delta C_{\text{C}}+C_{1}+C_{2}+J C_{\text{NL}}\left(v_{m,n}^{\left(3\right)}\right)\right]
\frac{d^{2}V_{m,n}^{(3)}}{dt^{2}} \nonumber \\
&&-\left(\frac{1}{\Delta L_{\text{C}}}+\frac{1}{L_{1}}+\frac{1}{L_{2}}\right)V_{m,n}^{\left(3\right)}  \nonumber \\
&=&\frac{V_{m,n}^{\left(3\right)}}{L_{\text{g}}}
+\left( C_{\text{g}}+J C_{\text{L}}\right) \frac{d^{2}V_{m,n}^{\left(3\right)}}{dt^{2}},  \label{circuit_eq_3} \\
&&C_{1}\frac{d^{2}V_{m,n}^{\left(2\right)}}{dt^{2}}
+C_{2}\frac{d^{2}V_{m+1,n}^{\left(2\right)}}{dt^{2}}
+C_{2}\frac{d^{2}V_{m,n-1}^{\left(1\right)}}{dt^{2}}
+C_{1}\frac{d^{2}V_{m,n}^{\left(1\right)}}{dt^{2}} \nonumber \\
&&-\left[\Delta C_{\text{D}}+2C_{1}+2C_{2}+J C_{\text{NL}}\left(v_{m,n}^{\left(4\right)}\right)\right]%
\frac{d^{2}V_{m,n}^{\left(4\right)}}{dt^{2}}
-\frac{V_{m,n}^{\left(4\right)}}{\Delta L_{\text{D}}} \nonumber \\
&=&\frac{V_{m,n}^{\left(4\right)}}{L_{\text{g}}}
+\left( C_{\text{g}}+J C_{\text{L}}\right) \frac{d^{2}V_{m,n}^{\left(4\right)}}{dt^{2}},  \label{circuit_eq_4}
\end{eqnarray}%
where the linear and nonlinear components of the diode capacitance have been
separated. These equations describe the time-dependent voltage distribution
across the circuit lattice.

The voltages at each circuit node can be expressed as $V_{m,n}^{(1,2,3,4)}=%
\frac{1}{2}v_{m,n}^{(1,2,3,4)}e^{-\mathrm{i}\omega _{0}t}+\text{c.c.},$ from
which we can obtain the first and second derivatives. Due to relations $C_{%
\text{g}}+J C_{\text{L}}\gg C_{1,2},C_{\text{NL}},\Delta C_{\text{A},%
\text{B},\text{C},\text{D}}$ and $L_{\text{g}}\ll L_{1,2}$, we apply the
slowly-varying-envelope approximation. In Eqs. (\ref{circuit_eq_1})-(\ref%
{circuit_eq_4}), we use
\begin{equation}
\frac{d^{2}V_{m,n}^{(1,2,3,4)}}{dt^{2}}=-\mathrm{i}\omega _{0}\frac{%
dv_{m,n}^{(1,2,3,4)}}{dt}e^{-\mathrm{i}\omega _{0}t}-\frac{\omega _{0}^{2}}{2%
}v_{m,n}^{(1,2,3,4)}e^{-\mathrm{i}\omega _{0}t}+\text{c.c.}
\end{equation}%
for the terms on the right-hand side. For the other terms, we instead use
\begin{equation}
\frac{d^{2}V_{m,n}^{(1,2,3,4)}}{dt^{2}}=-\frac{\omega _{0}^{2}}{2}%
v_{m,n}^{(1,2,3,4)}e^{-\mathrm{i}\omega _{0}t}+\text{c.c.}
\end{equation}%
Thus, Eqs. (\ref{circuit_eq_1})-(\ref{circuit_eq_4}) are reduced to
\begin{eqnarray}
\mathrm{i}\frac{dv_{m,n}^{(1)}}{dt} &=&
\left[-\frac{2C_{1}+2C_{2}+\Delta C_{\text{A}}+J C_{\text{NL}}\left(v_{m,n}^{\left(1\right)}\right)}{2(C_{\text{g}}+J C_{\text{L}})}
+\frac{L_{\text{g}}}{2}\frac{1}{\Delta L_{\text{A}}}\right]\omega _{0}v_{m,n}^{(1)}  \nonumber \\
&&+\frac{C_{1}}{2(C_{\text{g}}+J C_{\text{L}})}\omega _{0}v_{m,n}^{\left(3\right)}
+\frac{C_{2}}{2(C_{\text{g}}+J C_{\text{L}})}\omega _{0}v_{m+1,n}^{\left(3\right)}  \nonumber \\
&&+\frac{C_{1}}{2(C_{\text{g}}+J C_{\text{L}})}\omega _{0}v_{m,n}^{\left(4\right)}
+\frac{C_{2}}{2(C_{\text{g}}+J C_{\text{L}})}\omega _{0}v_{m,n+1}^{\left(4\right)}, \\
\mathrm{i}\frac{dv_{m,n}^{(2)}}{dt} &=&
\left[-\frac{C_{1}+C_{2}+\Delta C_{\text{B}}+J C_{\text{NL}}\left(v_{m,n}^{\left(2\right)}\right)}{2(C_{\text{g}}+J C_{\text{L}})} \right]\omega _{0}v_{m,n}^{(2)}  \nonumber \\
&&+\left[ \frac{L_{\text{g}}}{2}(\frac{1}{L_{1}}+\frac{1}{L_{2}}+\frac{1}{\Delta L_{%
\text{B}}})\right]\omega _{0}v_{m,n}^{(2)} \nonumber \\
&&-\frac{L_{\text{g}}}{2L_{1}}\omega_{0}v_{m,n}^{(3)}
-\frac{L_{\text{g}}}{2L_{2}}\omega _{0}v_{m,n-1}^{(3)} \nonumber \\
&&+\frac{C_{1}}{2(C_{\text{g}}+J C_{\text{L}})}\omega _{0}v_{m,n}^{\left(4\right)}
+\frac{C_{2}}{2(C_{\text{g}}+J C_{\text{L}})}\omega _{0}v_{m-1,n}^{\left(4\right)},
\end{eqnarray}
\begin{eqnarray}
\mathrm{i}\frac{dv_{m,n}^{(3)}}{dt} &=&
\left[-\frac{C_{1}+C_{2}+\Delta C_{\text{C}%
}+J C_{\text{NL}}\left(v_{m,n}^{\left(3\right)}\right)}{2(C_{\text{g}}+J C_{\text{L}})} \right]\omega _{0}v_{m,n}^{\left(3\right)}  \nonumber \\
&&+\left[\frac{L_{\text{g}}}{2}\left(\frac{1}{L_{1}}+\frac{1}{L_{2}}+\frac{1}{\Delta L_{\text{C}}}\right)\right]
\omega _{0}v_{m,n}^{\left(3\right)}  \nonumber \\
&&-\frac{L_{\text{g}}}{2L_{1}}\omega_{0}v_{m,n}^{\left(2\right)}
-\frac{L_{\text{g}}}{2L_{2}}\omega _{0}v_{m,n+1}^{(2)} \nonumber \\
&&+\frac{C_{1}}{2(C_{\text{g}}+J C_{\text{L}})}\omega _{0}v_{m,n}^{(1)}
+\frac{C_{2}}{2(C_{\text{g}}+J C_{\text{L}})}\omega _{0}v_{m-1,n}^{(1)}, \\
\mathrm{i}\frac{dv_{m,n}^{(4)}}{dt} &=&
\left[-\frac{2C_{1}+2C_{2}+\Delta C_{\text{%
D}}+J C_{\text{NL}}\left(v_{m,n}^{(4)}\right)}{2(C_{\text{g}}+J C_{\text{L}})}+%
\frac{L_{\text{g}}}{2}\frac{1}{\Delta L_{\text{D}}}\right]\omega _{0}v_{m,n}^{(4)}  \nonumber \\
&&+\frac{C_{1}}{2(C_{\text{g}}+J C_{\text{L}})}\omega _{0}v_{m,n}^{(1)}
+\frac{C_{2}}{2(C_{\text{g}}+J C_{\text{L}})}\omega _{0}v_{m,n-1}^{(1)} \nonumber \\
&&+\frac{C_{1}}{2(C_{\text{g}}+J C_{\text{L}})}\omega _{0}v_{m,n}^{(2)}+%
\frac{C_{2}}{2(C_{\text{g}}+J C_{\text{L}})}\omega _{0}v_{m+1,n}^{(2)}.
\end{eqnarray}%
We define $v_{m,n}^{(1,2,3,4)}(t)=V_{m,n}^{(1,2,3,4)}(t)\exp \left(\mathrm{i}%
\omega _{0}t\right)$, and, for the simplicity's sake, we also define $T=\omega
_{n}t$, where $\omega _{n}$ is the normalized frequency. Then, the equations
can be naturally written as
\begin{eqnarray}
\mathrm{i}\frac{dV_{m,n}^{(1)}}{dT} &=&\delta _{1}V_{m,n}^{(1)}+\gamma
_{C}V_{m,n}^{(3)}+\lambda _{C}V_{m+1,n}^{(3)}+\gamma
_{C}V_{m,n}^{(4)}+\lambda _{C}V_{m,n+1}^{(4)}  \nonumber \\
&&+g\left( V_{m,n}^{\left( 1\right) }\right) V_{m,n}^{(1)},  \label{eq1} \\
\mathrm{i}\frac{dV_{m,n}^{(2)}}{dT} &=&\delta _{2}V_{m,n}^{(2)}-\gamma
_{L}V_{m,n}^{(3)}-\lambda _{L}V_{m,n-1}^{(3)}+\gamma
_{C}V_{m,n}^{(4)}+\lambda _{C}V_{m-1,n}^{(4)}  \nonumber \\
&&+g\left( V_{m,n}^{\left( 2\right) }\right) V_{m,n}^{(2)},  \label{eq2} \\
\mathrm{i}\frac{dV_{m,n}^{(3)}}{dT} &=&\delta _{3}V_{m,n}^{(3)}-\gamma
_{L}V_{m,n}^{(2)}-\lambda _{L}V_{m,n+1}^{(2)}+\gamma
_{C}V_{m,n}^{(1)}+\lambda _{C}V_{m-1,n}^{(1)}  \nonumber \\
&&+g\left( V_{m,n}^{\left( 3\right) }\right) V_{m,n}^{(3)},  \label{eq3} \\
\mathrm{i}\frac{dV_{m,n}^{(4)}}{dT} &=&\delta _{4}V_{m,n}^{(4)}+\gamma
_{C}V_{m,n}^{(1)}+\lambda _{C}V_{m,n-1}^{(1)}+\gamma
_{C}V_{m,n}^{(2)}+\lambda _{C}V_{m+1,n}^{(2)}  \nonumber \\
&&+g\left( V_{m,n}^{\left( 4\right) }\right) V_{m,n}^{(4)},  \label{eq4}
\end{eqnarray}%
where
\begin{eqnarray}
\delta _{1} &=& \frac{\omega _{0}}{\omega _{n}}
-\frac{2C_{1}+2C_{2}+\Delta C_{\text{A}}}{2\left(C_{\text{g}}+J C_{\text{L}}\right)} \frac{\omega _{0}}{\omega _{n}}
+\frac{L_{\text{g}}}{2}\frac{1}{\Delta L_{\text{A}}} \frac{\omega _{0}}{\omega _{n}},  \label{delta_1} \\
\delta _{2}&=& \frac{\omega _{0}}{\omega _{n}}
-\frac{C_{1}+C_{2}+\Delta C_{\text{B}}}{2\left(C_{\text{g}}+JC_{\text{L}}\right)} \frac{\omega _{0}}{\omega _{n}} \nonumber \\
&&+\frac{L_{\text{g}}}{2}\left(\frac{1}{L_{1}}+\frac{1}{L_{2}}+\frac{1%
}{\Delta L_{\text{B}}}\right) \frac{\omega _{0}}{\omega _{n}},  \label{delta_2} \\
\delta _{3} &=&\frac{\omega _{0}}{\omega _{n}}
-\frac{C_{1}+C_{2}+\Delta C_{\text{C}}}{2\left(C_{\text{g}}+J C_{\text{L}}\right)}  \frac{\omega _{0}}{\omega _{n}} \nonumber \\
&&+\frac{L_{\text{g}}}{2}\left(\frac{1}{L_{1}}+\frac{1}{L_{2}}+\frac{1%
}{\Delta L_{\text{C}}}\right)  \frac{\omega _{0}}{\omega _{n}},  \label{delta_3} \\
\delta _{4} &=&\frac{\omega_{0}}{\omega _{n}}
-\frac{2C_{1}+2C_{2}+\Delta C_{\text{D}}}{2\left(C_{\text{g}%
}+J C_{\text{L}}\right)}\frac{\omega_{0}}{\omega _{n}}
+\frac{L_{\text{g}}}{2\Delta L_{\text{D}}} \frac{\omega
_{0}}{\omega _{n}},  \label{delta_4}
\end{eqnarray}%
represent the constant onsite energies,
\begin{eqnarray}
\gamma _{C} &=&\frac{C_{1}}{2(C_{\text{g}}+J C_{\text{L}})}\frac{\omega
_{0}}{\omega _{n}},  \label{gamma_C} \\
\lambda _{C} &=&\frac{C_{2}}{2(C_{\text{g}}+J C_{\text{L}})}\frac{\omega
_{0}}{\omega _{n}}, \\
\gamma _{L} &=&\frac{L_{\text{g}}}{2L_{1}}\frac{\omega _{0}}{\omega _{n}},
\label{gamma_L} \\
\lambda _{L} &=&\frac{L_{\text{g}}}{2L_{2}}\frac{\omega _{0}}{\omega _{n}},
\label{lambda_L}
\end{eqnarray}
represent the hopping strengths, and
\begin{equation}
g\left( V_{m,n}^{\left( 1,2,3,4\right) }\right) =-\frac{J C_{\text{NL}%
}\left( V_{m,n}^{\left( 1,2,3,4\right) }\right) }{2(C_{\text{g}}+J C_{%
\text{L}})}\frac{\omega _{0}}{\omega _{n}}
\end{equation}%
determines the amplitude-dependent onsite energies. These equations govern
the evolution in the circuit lattice.

To emulate the NLQTI, the coefficients must satisfy constraints $\delta
_{1}=\delta _{2}=\delta _{3}=\delta _{4}\equiv \delta $, $\gamma _{C}=\gamma
_{L}\equiv \gamma $, and $\lambda _{C}=\lambda _{L}\equiv \lambda $. To find
the most basic NLQTI states, we also set $\Delta C_{\text{A}}=\Delta C_{%
\text{D}}=0$, $\Delta C_{\text{B}}=\Delta C_{\text{C}}=C_{1}+C_{2}$, $\frac{1%
}{\Delta L_{\text{B}}}=\frac{1}{\Delta L_{\text{C}}}=0$, and $\frac{1}{%
\Delta L_{\text{A}}}=\frac{1}{\Delta L_{\text{D}}}=\frac{1}{L_{1}}+\frac{1}{%
L_{2}}$, which leads to the following relations between the circuit
parameters:
\begin{eqnarray}
C_{\text{A}} &=&C_{\text{D}}=C_{\text{g}},  \label{C_AD} \\
L_{\text{B}} &=&L_{\text{C}}=L_{\text{g}}, \\
C_{\text{B}} &=&C_{\text{C}}=C_{\text{g}}+C_{1}+C_{2}, \\
\frac{1}{L_{\text{A}}} &=&\frac{1}{L_{\text{D}}}=\frac{1}{L_{\text{g}}}+%
\frac{1}{L_{1}}+\frac{1}{L_{2}}, \\
C_{1} &=&\frac{L_{\text{g}}(C_{\text{g}}+J C_{\text{L}})}{L_{1}}, \\
C_{2} &=&\frac{L_{\text{g}}(C_{\text{g}}+J C_{\text{L}})}{L_{2}}.
\label{C2}
\end{eqnarray}

\section{The consideration on the linear quadrupole topological
insulator (QTI)\label{app_linear_QI}}

In this section, we demonstrate how the model of the NLQTI can be reduced to
its linear counterpart, which corresponds exactly to the seminal lattice
model for the linear QTI \cite{science357-61}.

In the linear limit, where $C_{\text{NL}}=0$, Eqs. (\ref{eq1})-(\ref{eq4})
reduce to the following equations:
\begin{eqnarray}
\mathrm{i}\frac{dV_{m,n}^{(1)}}{dT} &=&\delta _{1}V_{m,n}^{(1)}+\gamma
_{C}V_{m,n}^{(3)}+\lambda _{C}V_{m+1,n}^{(3)}+\gamma
_{C}V_{m,n}^{(4)} \nonumber \\
&&+\lambda _{C}V_{m,n+1}^{(4)}, \\
\mathrm{i}\frac{dV_{m,n}^{(2)}}{dT} &=&\delta _{2}V_{m,n}^{(2)}-\gamma
_{L}V_{m,n}^{(3)}-\lambda _{L}V_{m,n-1}^{(3)}+\gamma
_{C}V_{m,n}^{(4)} \nonumber \\
&&+\lambda _{C}V_{m-1,n}^{(4)}, \\
\mathrm{i}\frac{dV_{m,n}^{(3)}}{dT} &=&\delta _{3}V_{m,n}^{(3)}-\gamma
_{L}V_{m,n}^{(2)}-\lambda _{L}V_{m,n+1}^{(2)}+\gamma
_{C}V_{m,n}^{(1)} \nonumber \\
&&+\lambda _{C}V_{m-1,n}^{(1)}, \\
\mathrm{i}\frac{dV_{m,n}^{(4)}}{dT} &=&\delta _{4}V_{m,n}^{(4)}+\gamma
_{C}V_{m,n}^{(1)}+\lambda _{C}V_{m,n-1}^{(1)}+\gamma
_{C}V_{m,n}^{(2)} \nonumber \\
&&+\lambda _{C}V_{m+1,n}^{(2)},
\end{eqnarray}%
where $\delta _{1,2,3,4}$, $\gamma _{C,L}$, and $\lambda _{C,L}$ retain the
same form as defined in Eqs. (\ref{delta_1})-(\ref{lambda_L}). To establish
the correspondence between our circuit model and the tight-binding one
proposed in Ref. \cite{science357-61}, the coefficients are subject to
constraints $\delta _{1}=\delta _{2}=\delta _{3}=\delta _{4}\equiv \delta $,
$\gamma _{C}=\gamma _{L}\equiv \gamma $, and $\lambda _{C}=\lambda
_{L}\equiv \lambda $. We define the circuit parameters in the same way as
those in Eqs. (\ref{C_AD})-(\ref{C2}), which reduce to
\begin{eqnarray}
\mathrm{i}\frac{dV_{m,n}^{(1)}}{dT} &=&\delta V_{m,n}^{(1)}+\gamma
V_{m,n}^{(3)}+\lambda V_{m+1,n}^{(3)}+\gamma V_{m,n}^{(4)}+\lambda
V_{m,n+1}^{(4)}, \\
\mathrm{i}\frac{dV_{m,n}^{(2)}}{dT} &=&\delta V_{m,n}^{(2)}-\gamma
V_{m,n}^{(3)}-\lambda V_{m,n-1}^{(3)}+\gamma V_{m,n}^{(4)}+\lambda
V_{m-1,n}^{(4)}, \\
\mathrm{i}\frac{dV_{m,n}^{(3)}}{dT} &=&\delta V_{m,n}^{(3)}-\gamma
V_{m,n}^{(2)}-\lambda V_{m,n+1}^{(2)}+\gamma V_{m,n}^{(1)}+\lambda
V_{m-1,n}^{(1)}, \\
\mathrm{i}\frac{dV_{m,n}^{(4)}}{dT} &=&\delta V_{m,n}^{(4)}+\gamma
V_{m,n}^{(1)}+\lambda V_{m,n-1}^{(1)}+\gamma V_{m,n}^{(2)}+\lambda
V_{m+1,n}^{(2)}.
\end{eqnarray}%
For the linear QTI with periodic boundary conditions in the $x$ and $y$
directions, the respective Hamiltonian in the $k$-space can be written as
\begin{eqnarray}
H&=&\left( \gamma +\lambda \cos k_{x}\right) \Gamma _{4}+\lambda \sin
k_{x}\Gamma _{3}+\left( \gamma +\lambda \cos k_{y}\right) \Gamma
_{2} \nonumber \\
&&+\lambda \sin k_{y}\Gamma _{1} +\delta \Gamma _{0},  \label{Hamiltonian}
\end{eqnarray}%
where $\Gamma _{0}=\tau _{0}\sigma _{0}$, $\Gamma _{1}=-\tau _{2}\sigma _{1}$%
, $\Gamma _{2}=-\tau _{2}\sigma _{2}$, $\Gamma _{3}=-\tau _{2}\sigma _{3}$,
and $\Gamma _{4}=\tau _{1}\sigma _{0}$, with $\tau $ and $\sigma $ being two
sets of the Pauli matrices ($\tau _{0}$ and $\sigma _{0}$ stand for the unit
matrix). The corresponding eigenfrequencies are given by
\begin{equation}
\bar{\omega}=\delta \pm \sqrt{2\gamma ^{2}+2\lambda ^{2}+2\gamma \lambda
\left( \cos k_{x}+\cos k_{y}\right) }.  \label{omega}
\end{equation}%
Both the upper and lower energy bands are twofold degenerate. A frequency
gap (equivalent to the energy gap in Ref. \cite{science357-61}) exists
unless $\left\vert \gamma /\lambda \right\vert =1$. Accordingly, the phase
transition occurs at $\gamma /\lambda =1$ ($\gamma /\lambda =-1$), with the
corresponding bulk frequency gap closing at the $\mathbf{M}=\left( \pi ,\pi
\right) $ ($\mathbf{\Gamma }=\left( 0,0\right) $) point of the Brillouin
zone \cite{science357-61}. When $\left\vert \gamma /\lambda \right\vert \neq
1$, the size of the bandgap, according to Eq. (\ref{omega}), is $2\sqrt{2}%
\left\vert \gamma -\lambda \right\vert $ when $\gamma $ and $\lambda $ have
the same sign, and by $2\sqrt{2}\left\vert \gamma +\lambda \right\vert $
when $\gamma $ and $\lambda $ have different signs. Finally, the bands
become flat when $\gamma =0$ or $\lambda =0$.

The linear QTI exhibits several symmetry properties \cite{science357-61}. First, the system
maintains reflection symmetries, as the Hamiltonian satisfies the relation
\begin{equation}
m_{j}H\left(k\right)m_{j}^{\dagger }=H\left(M_{j}k\right),
\end{equation}%
for $j=x,y$, where $m_{x}=\tau _{1}\sigma _{3}$, $m_{y}=\tau _{1}\sigma _{1}$%
, with $M_{x}\left(k_{x},k_{y}\right)=\left(-k_{x},k_{y}\right)$ and $%
M_{y}\left(k_{x},k_{y}\right)=\left(k_{x},-k_{y}\right)$. Both reflection symmetries hold for the
present lattice model. Note that matrices $m_{x}$ and $m_{y}$ do not
commute; instead, they anticommute, so that $\{m_{x},m_{y}\}=0$ in the
present model. Second, the model also maintains the time-reversal symmetry,
with $T=K$ as the time-reversal operator. Here, $K$ is the
complex-conjugation operator. In the coordinate space, this symmetry
requires that
\begin{equation}
THT^{-1}=H,
\end{equation}%
where $T=T^{-1}=K$. The latter condition reduces to $H^{\ast }=H$, meaning
that the Hamiltonian is real in the coordinate space, which is indeed the
case. In the $k$-space, the time-reversal symmetry demands that
\begin{equation}
TH(\mathbf{k})T^{-1}=H(-\mathbf{k}),
\end{equation}%
which simplifies to $H^{\ast }(\mathbf{k})=H(-\mathbf{k})$. This condition
is satisfied by Hamiltonian (\ref{Hamiltonian}). Third, the model maintains
the charge-conjugation symmetry, with $C=\Gamma _{0}K$ acting as the
charge-conjugation operator. In the $k$-space, this symmetry requires that
\begin{equation}
CH(\mathbf{k})C^{-1}=-H(-\mathbf{k}),
\end{equation}%
where $C=C^{-1}=\Gamma _{0}K$. This condition is also satisfied by
Hamiltonian (\ref{Hamiltonian}). Fourth,
the model preserves the $C_{4}$ symmetry up to a gauge transformation.
In the $k$-space, this symmetry requires that
\begin{equation}
r_{4}H(\mathbf{k})r_{4}^{\dagger }=H(R_{4}\mathbf{k}),
\end{equation}%
with
\begin{equation}
r_{4}=\frac{\tau _{1}+i\tau _{2}}{2}\otimes \sigma _{0}-\frac{\tau
_{1}-i\tau _{2}}{2}\otimes i\sigma _{2},
\end{equation}%
and $R_{4}$ representing the rotation of the momentum by $\pi /2$. This
symmetry is clearly satisfied, as
\begin{equation}
r_{4}H(k_{x},k_{y})r_{4}^{\dagger }=H(k_{y},-k_{x}).
\end{equation}%
Fifth, considering that the chiral symmetry is defined by operator $C=\tau
_{3}\sigma _{0}$, and taking into regard that the terms representing the
onsite energies in the linear lattice can be eliminated by a gauge
transformation, the model also maintains the chiral symmetry, with the
Hamiltonian satisfying relation%
\begin{equation}
CH(\mathbf{k})C^{\dagger }=-H(\mathbf{k}).
\end{equation}%
Finally, the model exhibits the inversion symmetry. As the inversion
symmetry $I$ is related to the mirror symmetries by $I=m_{y}m_{x}$, the
present model is indeed an inversion-symmetric one.

\begin{figure}[t]
\centering
\includegraphics[width=4.5cm]{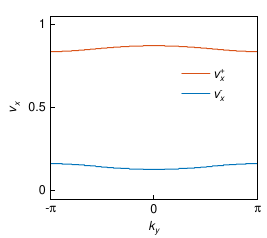}
\caption{\textbf{The Wannier bands $\protect\nu _{x}^{\pm }(k_{y})$
calculated from the the Bloch Hamiltonian by means of the Wilson loop
diagonalization.} The circuit parameters are $C_{\text{L}}=4.3~\text{nF}$, $%
v_{0}=1.7$, $M=0.3$, $J =2$, $L_{\text{g}}=1~\mathrm{\protect\mu H%
}$, $L_{1}=150~\mathrm{\protect\mu H}$, and $L_{2}=15~\mathrm{\protect\mu H}$%
, with the values of the other circuit elements determined accordingly.}
\label{fig_wannier_band}
\end{figure}

Due to the presence of these symmetries, the quadrupole moment $q_{xy}$ is
effectively quantized and can be calculated using the method of nested
Wilson loops \cite{science357-61}. We first calculate the Wilson loop in the
$x$ direction, denoted as $\mathcal{W}_{x,k}$, where $k=(k_{x},k_{y})$
represents the starting point of the loop. For simplicity, we denote the
Bloch functions at $k$ as $\left\vert u_{k}^{n}\right\rangle $, where $n=1$
or $2$ indicates the band index for the two bands below the bandgap. The
components of the Bloch functions are denoted as $\left[ u_{k}^{n}\right]
^{\alpha }$ with $\alpha =1,\ldots ,4$. We define $\left[ F_{x,k}\right]
^{mn}=\left\langle u_{k+\Delta k_{x}}^{m}\middle|u_{k}^{n}\right\rangle $,
where $\Delta k_{x}=\left( \frac{2\pi }{N_{x}},0\right) $ and $N_{x}$ is the
number of lattice sites along the $x$ direction. Therefore, we have
\begin{equation}
\mathcal{W}_{x,k}=F_{x,k+\left( N_{x}-1\right) \Delta k_{x}}\cdots
F_{x,k+\Delta k_{x}}F_{x,k}.
\end{equation}%
Based on the Wilson loop operator $\mathcal{W}_{x,k}$, a Wannier Hamiltonian
$H_{\mathcal{W}_{x}(k)}$ can be defined via
\begin{equation}
\mathcal{W}_{x,k}=e^{\mathrm{i}H_{\mathcal{W}_{x}(k)}}.
\end{equation}%
Considering that $H_{\mathcal{W}_{x}(k)}$ has eigenvalues $2\pi
v_{x}^{j}(k_{y})$ for $j=+,-$, where $v_{x}^{j}(k_{y})$ are the Wannier
centers in the $x$ direction, the Wilson loop can be diagonalized under
fully periodic boundary conditions as
\begin{equation}
\mathcal{W}_{x,k}=\sum_{j=\pm }\left\vert v_{x,k}^{j}\right\rangle e^{2\pi
\mathrm{i}v_{x}^{j}(k_{y})}\left\langle v_{x,k}^{j}\right\vert .
\end{equation}%
Here, the eigenstates $\left\vert v_{x,k}^{j}\right\rangle $ for $j=+,-$
have components $\left[ v_{x,k}^{j}\right] ^{n}$ with $n=1,2$. Due to the $%
x\rightarrow -x$ reflection symmetry, we have $%
v_{x}^{-}(k_{y})=-v_{x}^{+}(k_{y})$ mod $1$. For the definiteness' sake, we
choose $v_{x}^{-}(k_{y})\in \lbrack 0,0.5]$. We set the circuit parameters as $C_{\text{L}}=4.3~\text{nF}$%
, $v_{0}=1.7$, $M=0.3$, $J =2$, $L_{\text{g}}=1~\mathrm{\mu H}$, $%
L_{1}=150~\mathrm{\mu H}$, and $L_{2}=15~\mathrm{\mu H}$, with the values of
the other circuit elements determined accordingly. As shown in Fig. \ref%
{fig_wannier_band}, this operation splits the original degenerate bands into
two gapped Wannier sectors, labeled as $v_{x}^{-}$ and $v_{x}^{+}$. Since
they are gapped, each of the Wannier sectors carries its own set of
topological invariants.

We define the Wannier band subspaces as
\begin{equation}
\left\vert w_{x,k}^{\pm }\right\rangle =\sum_{n=1,2}\left\vert
u_{k}^{n}\right\rangle \left[ v_{x,k}^{\pm }\right] ^{n}.
\end{equation}%
Using these new subspaces, we further define $F_{y,k}^{\pm }=\left\langle
w_{x,k+\Delta k_{y}}^{\pm }\middle|w_{x,k}^{\pm }\right\rangle $, where $%
\Delta k_{y}=\left( 0,\frac{2\pi }{N_{y}}\right) $ and $N_{y}$ is the number
of lattice sites along the $y$ direction. The nested Wilson loops along $%
k_{y}$ in the Wannier bands $v_{x}^{\pm }$ are given by
\begin{equation}
\mathcal{\tilde{W}}_{y,k_{x}}^{\pm }=F_{y,k+\left( N_{y}-1\right) \Delta
k_{y}}^{\pm }\cdots F_{y,k+\Delta k_{y}}^{\pm }F_{y,k}^{\pm }.
\end{equation}%
The associated Wannier-sector polarizations are
\begin{eqnarray}
p_{y}^{v_{x}^{\pm }} &=&-\frac{\mathrm{i}}{2\pi }\frac{1}{N_{x}}%
\sum_{k_{x}}\log \left[ \mathcal{\tilde{W}}_{y,k_{x}}^{\pm }\right] \\
&=&\left\{
\begin{array}{ll}
0, & \gamma /\lambda >1, \\
\frac{1}{2}, & \ \gamma /\lambda <1,%
\end{array}%
\right.
\end{eqnarray}%
where $v_{x}^{\pm }$ refer to the Wannier sectors. The polarization value $%
\frac{1}{2}$ in the $y$ direction indicates that the Wannier Hamiltonian of
the $x$ edge, $H_{\mathcal{W}_{x}(k)}$, represents a topological insulator
when $\gamma /\lambda <1$. Similarly, we can compute the Wannier-sector
polarization $p_{x}^{v_{y}^{\pm }}$. Finally, the quadrupole moment $%
q_{xy}^{\pm }$ is defined as
\begin{equation}
q_{xy}^{\pm }=2p_{x}^{v_{y}^{\pm }}p_{y}^{v_{x}^{\pm }}.
\end{equation}%
Due to $q_{xy}^{+}\equiv q_{xy}^{-}$, we denote the quadrupole moment as $%
q_{xy}$. This quadrupole moment equals $\frac{1}{2}$ for $\gamma <\lambda $,
and vanishes for $\gamma >\lambda $. Nonzero $q_{xy}$ implies that the
linear lattice is a topologically nontrivial one.

\begin{figure}[t]
\centering
\includegraphics[width=8.6cm]{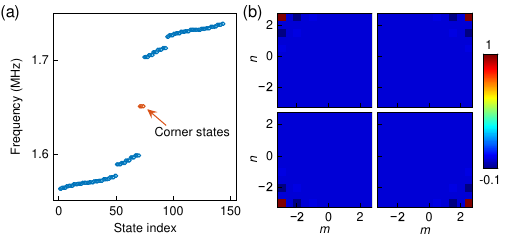}
\caption{\textbf{Eigenfrequencies and topological corner states of the LQTI.}
(\textbf{a}) Eigenfrequencies of the LQTI. The four eigenfrequencies near
the center (marked by red dots) correspond to the topological corner states.
(\textbf{b}) Voltage distributions of the four topological corner states, normalized to their respective maximum values.}
\label{fig_linear}
\end{figure}

We next investigate a linear QTI with open boundary conditions in the $x$
and $y$ directions, based on the system consisting of $N_{x}$ and $N_{y}$
unit cells along the $x$ and $y$ axes, respectively. By expressing the node
voltages as $V_{m,n}^{(1,2,3,4)}=\psi _{m,n}^{(1,2,3,4)}\exp \left(-\mathrm{i}%
\bar{\omega}T\right)$, the governing equations can be cast in a matrix form, from
which the eigenfrequencies and eigenstates of the linear QTI are obtained.
We consider a circuit lattice with $N_{x}=N_{y}=6$. The parameters of the
common-cathode diodes are $C_{\text{L}}\ =4.3~\text{nF}$, $v_{0}=1.7$, and $%
M=0.3$, as determined by fitting to experimental measurements. To enhance
the nonlinear response, two common-cathode diodes are used, corresponding to
$J =2$, where $J $ is the number of diodes. The grounded inductor is
fixed by setting $L_{\text{g}}=1~\mathrm{\mu }\text{H}$, and the grounded
capacitor is absent, $C_{\text{g}}=0$. For the system to exhibit the
topological quadrupole behavior, the intracell hopping must be weaker than
its intercell counterpart, i.e., $C_{1}<C_{2}$ and $L_{1}>L_{2}$. To satisfy
the slowly-varying-envelope approximation and ensure strong localization of
the corner states, the coupling inductors are chosen with $L_{1}=150~\mathrm{%
\mu }\text{H}$ and $L_{2}=15~\mathrm{\mu }\text{H}$. Based on these values,
the remaining circuit elements can be determined accordingly. Figure \ref%
{fig_linear}(a) presents the eigenfrequency spectrum of the linear QTI, with
the four eigenfrequencies near the center (marked by red dots) corresponding
to the topological corner states. Generally speaking, in the linear QTI
consisting of $N_{x}\times N_{y}$ unit cells, there are four corner states, $%
2\left(N_{x}-2\right)+2\left(N_{y}-2\right)$ edge states, and $4\left(N_{x}-1\right)\left(N_{y}-1\right)$ bulk states.
The voltage distributions of the four topological corner states, normalized to their respective maximum values,
are illustrated in Fig.~\ref{fig_linear}(b).

\section{Nonlinear topological corner states and topologically trivial
corner solitons\label{app_corner}}

In this section, we first discuss the localization of nonlinear topological
corner states and topologically trivial corner solitons. Next, we examine
the topological properties of these two types of states. Finally, we address
the dynamical stability or instability of the nonlinear topological corner
states and topologically trivial corner solitons.

\subsection{The localization\label{app_localization}}

We consider the NLQTI with open boundary conditions in the $x$ and $y$
directions, neglecting dissipation in the circuit lattice. By substituting $%
V_{m,n}^{(1,2,3,4)}=\psi _{m,n}^{(1,2,3,4)}\exp \left( -\mathrm{i}\bar{\omega%
}T\right) $ into Eqs.~(\ref{eq1})-(\ref{eq4}), the resulting equations are
solved using the Newton's method. For each value of $\bar{\omega}$, we
adopted initial guesses for the voltage distributions. Once the solution is
obtained at a given $\bar{\omega}$, solutions at other frequencies are
derived iteratively. When searching for the nonlinear corner states, we
focused on those residing in the upper-right corner of the lattice, as the
four linear topological corner states (denoted by the red dots in Fig. \ref%
{fig_linear}(a)) are equivalent to one another.

\begin{figure}[t]
\centering
\includegraphics[width=4.3cm]{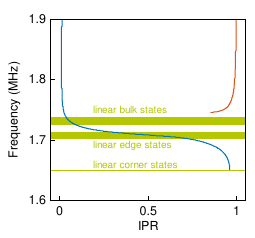}
\caption{\textbf{The localization of the nonlinear topological corner states
and topologically trivial corner solitons.} The blue and red curves
represent the IPRs of the nonlinear corner states and corner solitons,
respectively.}
\label{fig_IPR_corner}
\end{figure}

After obtaining the frequency spectra, we characterized the state
localization by their IPR, as per Eq. (\ref{IPR}) in the main text.
When the state is strongly localized at a single lattice site, we
have $\text{IPR}=1$, whereas for a state that is uniformly distributed
across the entire lattice, IPR$~=\left( \allowbreak 4N_{x}N_{y}\right) ^{-1}$%
. Figure~\ref{fig_IPR_corner} shows the IPRs of the nonlinear topological
corner states and topologically trivial corner solitons. We first focus on
the nonlinear corner states, which bifurcate from the linear corner mode.
When the frequency is below that of the linear edge states, the nonlinear
corner states are mainly localized at the corner site. As the frequency
increases, the localization weakens. Upon entering the edge-mode band, the
nonlinear corner states delocalize and, accordingly, the IPR rapidly
decreases, due to the hybridization between the corner and edge states. The
further increase of the frequency makes the nonlinear corner state even more
delocalized, due to the hybridization with the bulk modes.

Next, we turn to the corner solitons. Being, matter-of-factly, conventional
topologically trivial solitons, they become more localized with the increase
of the frequency. In the strongly nonlinear regime, these solitons are
strongly localized at the corner site, resulting in $\text{IPR}=1$.

\subsection{Topological properties\label{app_topo_property}}

\begin{figure*}[t]
\centering
\includegraphics[width=17.2cm]{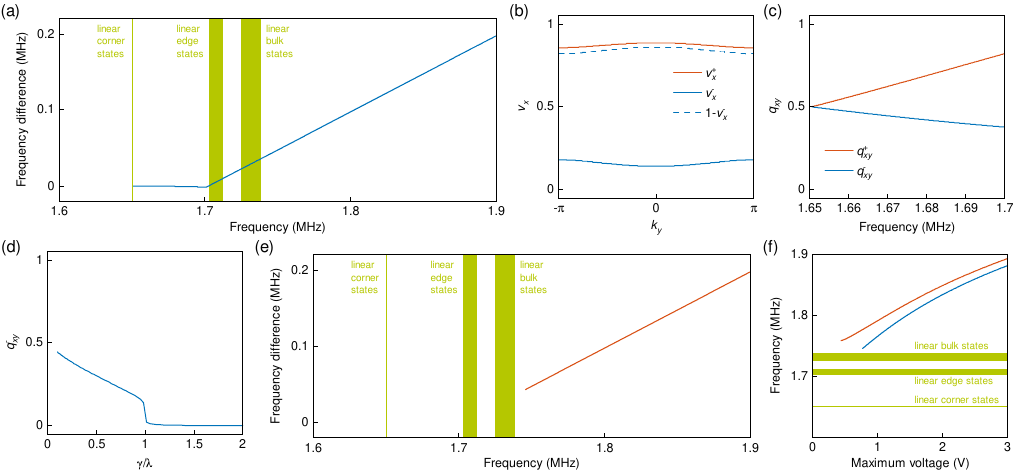}
\caption{\textbf{Topological properties of the NLQTI.} (\textbf{a}) The
difference between the eigenfrequencies of the nonlinear corner states and
those of the linear QTI, produced by Eqs. (\protect\ref{new_LQTI_1})-(%
\protect\ref{new_LQTI_4}). (\textbf{b}) The Wannier bands $\protect\nu %
_{x}^{\pm }(k_{y})$ calculated from the Bloch Hamiltonian derived from Eqs. (%
\protect\ref{new_LQTI_1})-(\protect\ref{new_LQTI_4}). For comparison, $%
1-v_{x}^{-}(k_{y})$ is also plotted as a dashed curve. The value of $g\left(
V_{2,2}^{(1)}\right) $ is determined by selecting the nonlinear corner state
at the frequency of $1.67\,\text{MHz}$. (\textbf{c}) The quadrupole moments $%
q_{xy}^{\pm }$, labeled with $\pm$ for their respective Wannier sectors,
at various frequencies of the nonlinear topological corner states. (\textbf{d%
}) The dependence of the quadrupole moment $q_{xy}^{-}$ on the dimerization
ratio $\protect\gamma /\protect\lambda $. The value of $g\left(
V_{2,2}^{(1)}\right) $, obtained by selecting the nonlinear corner state at
a frequency of $1.67\,\text{MHz}$, is used for calculations at different
dimerization ratios. (\textbf{e}) The difference between the
eigenfrequencies of the corner solitons and those of the LQTI governed by
Eqs. (\protect\ref{new_LQTI_1})-(\protect\ref{new_LQTI_4}). (\textbf{f}) The
dependence of the eigenfrequency of the corner solitons on the maximum
voltage in the lattice. The red and blue curves represent the corner
solitons calculated from Eqs. (\protect\ref{eq1})-(\protect\ref{eq4}) and
Eqs. (\protect\ref{NL_eq1})-(\protect\ref{NL_eq4}), respectively.}
\label{fig_topo}
\end{figure*}

As discussed in \ref{app_linear_QI}, the linear QTI satisfies
several symmetry operations and the quadrupole moment $q_{xy}$ is quantized
under these symmetries. In the nonlinear case, we consider three regimes:
the weakly nonlinear regime, the moderately nonlinear regime, and the
strongly nonlinear regime. The weakly nonlinear regime is defined as the
region where nonlinear corner states are spatially localized. The moderately
nonlinear regime is defined as the region where nonlinear corner states are
delocalized and topologically trivial corner solitons do not exist. The
strongly nonlinear regime is defined as the region where corner solitons
exist. We explore the topological properties of the NLQTI in what follows
below.

The onsite energy at a given lattice site of the NLQTI depends on the
voltage applied to that site. Strictly speaking, inhomogeneous voltage
distributions result in different onsite energies at various lattice sites,
leading to the absence of discrete translational symmetry in the circuit
lattice. However, in the weakly nonlinear regime, the nonlinear corner
states exhibit sublattice polarization \cite{SP18-208}, allowing the NLQTI to be
approximated as a linear QTI where the condition $\delta _{1}=\delta
_{2}=\delta _{3}=\delta _{4}$ does not hold. Specifically, the nonlinear
corner state located in the upper-right corner of the lattice is polarized
in sublattice $1$, where the voltage is primarily distributed. This
polarization allows us to construct a linear QTI as follows:
\begin{eqnarray}
\mathrm{i}\frac{dV_{m,n}^{(1)}}{dT} &=&\left[ \delta _{1}+g\left(
V_{2,2}^{\left( 1\right) }\right) \right] V_{m,n}^{(1)}+\gamma
_{C}V_{m,n}^{(3)}+\lambda _{C}V_{m+1,n}^{(3)} \nonumber \\&&
+\gamma _{C}V_{m,n}^{(4)}
+\lambda _{C}V_{m,n+1}^{(4)},  \label{new_LQTI_1} \\
\mathrm{i}\frac{dV_{m,n}^{(2)}}{dT} &=&\delta _{2}V_{m,n}^{(2)}-\gamma
_{L}V_{m,n}^{(3)}-\lambda _{L}V_{m,n-1}^{(3)}+\gamma
_{C}V_{m,n}^{(4)} \nonumber \\ &&
+\lambda _{C}V_{m-1,n}^{(4)},  \label{new_LQTI_2} \\
\mathrm{i}\frac{dV_{m,n}^{(3)}}{dT} &=&\delta _{3}V_{m,n}^{(3)}-\gamma
_{L}V_{m,n}^{(2)}-\lambda _{L}V_{m,n+1}^{(2)}+\gamma
_{C}V_{m,n}^{(1)} \nonumber \\ &&
+\lambda _{C}V_{m-1,n}^{(1)},  \label{new_LQTI_3} \\
\mathrm{i}\frac{dV_{m,n}^{(4)}}{dT} &=&\delta _{4}V_{m,n}^{(4)}+\gamma
_{C}V_{m,n}^{(1)}+\lambda _{C}V_{m,n-1}^{(1)}+\gamma
_{C}V_{m,n}^{(2)} \nonumber \\ &&
+\lambda _{C}V_{m+1,n}^{(2)},  \label{new_LQTI_4}
\end{eqnarray}%
where $V_{2,2}^{\left( 1\right) }$ represents the voltage at site $1$ of the
$\left( 2,2\right) $-th unit cell, i.e., the corner site of the circuit
lattice. To verify the validity of this approximation, we calculate the
eigenfrequency of the corner state in the linear QTI governed by Eqs. (\ref%
{new_LQTI_1})-(\ref{new_LQTI_4}) and compare it with the eigenfrequency of
the nonlinear corner state. As shown in Fig. \ref{fig_topo}(a), the
difference between the eigenfrequencies is nearly zero for the localized
nonlinear corner states. In contrast, when the nonlinear corner states
become delocalized, the frequency difference is significant, indicating that
the approximation is no longer applicable.

In Eqs. (\ref{new_LQTI_1})-(\ref{new_LQTI_4}), by considering the condition $%
\delta _{1}=\delta _{2}=\delta _{3}=\delta _{4}\equiv \delta $, we have $%
\delta _{1}+g\left( V_{2,2}^{(1)}\right) >\delta $. As a result of these
relations, both reflection symmetries $m_{x}$ and $m_{y}$ are broken.
Furthermore, the charge-conjugation symmetry, $C_{4}$ symmetry, chiral
symmetry, and inversion symmetry are all violated, only the time-reversal
symmetry being preserved. We then employ the method of nested Wilson loops
to calculate the quadrupole moments $q_{xy}^{\pm }$. As an example, we
determine the value of $g\left( V_{2,2}^{(1)}\right) $ by selecting the
nonlinear corner state at a frequency of $1.67\,\text{MHz}$. This value is
subsequently used to compute the Wannier centers by following the procedure
outlined in \ref{app_linear_QI}. We find that the two Wannier
sectors, $v_{x}^{-}$ and $v_{x}^{+}$, remain gapped; however, $%
v_{x}^{-}(k_{y})\neq -v_{x}^{+}(k_{y})\mod 1$ due to the violation of
reflection symmetry in the $x$ direction, as demonstrated in Fig. \ref%
{fig_topo}(b). For comparison, we also plot $1-v_{x}^{-}(k_{y})$ as a dashed
curve.

In the linear QTI phase, the Wannier sector polarizations $p_{y}^{v_{x}^{\pm
}}$ are quantized due to the presence of two anti-commuting reflection
symmetries $m_{x}$ and $m_{y}$. However, in the nonlinear case, these
reflection symmetries are violated, resulting in $p_{y}^{v_{x}^{\pm }}$ not
being quantized. Similarly, we find that the Wannier sector polarizations $%
p_{x}^{v_{y}^{\pm }}$ are also not quantized. Consequently, the quadrupole
moments $q_{xy}^{\pm }$, labeled with $\pm $ to denote the respective
Wannier sectors, do not equal the quantized value of $1/2$, as shown in Fig. %
\ref{fig_topo}(c). Note that $q_{xy}^{+}\neq q_{xy}^{-}\neq 1/2$ when the
frequency deviates from that of the linear topological corner state.

Furthermore, when we evaluate the dependence of the quadrupole moment $%
q_{xy}^{-}$ on the dimerization ratio $\gamma / \lambda$, we observe that $%
q_{xy}^{-}$ varies continuously, as illustrated in Fig. \ref{fig_topo}(d).
However, the quadrupole moment remains strictly positive when $\gamma <
\lambda$, in contrast to the nearly zero values observed when $\gamma >
\lambda$. In our calculations, we use the same value of $g\left(
V_{2,2}^{(1)} \right)$ for different dimerization ratios. Specifically, this
value is determined by selecting the nonlinear corner state at a frequency
of $1.67\, \text{MHz}$. Therefore, these results imply that in this weakly
nonlinear regime, the nonzero quadrupole invariant can still be used to
demonstrate that the corner states are topologically nontrivial ones,
exhibiting robustness against perturbations induced by onsite nonlinearity.

In the moderately nonlinear regime, as shown in Fig. \ref{fig_topo}(a), the
approximations for the onsite energies are no longer valid, and Eqs. (\ref%
{new_LQTI_1})-(\ref{new_LQTI_4}) are no longer applicable for characterizing
the topological properties of the NLQTI. The delocalized field distribution
leads to a breakdown of the periodicity in the circuit lattice.
Consequently, the definition of the quadrupole moments $q_{xy}^{\pm}$,
introduced in \ref{app_linear_QI}, cannot be applied in this
context. In fact, due to the competition between onsite perturbations and
intersite hoppings, localized states are absent in this moderately nonlinear
regime.

In the regime of strong nonlinearity, the approximations for the onsite
energies become invalid. As shown in Fig. \ref{fig_topo}(e), the difference
between the eigenfrequencies of the corner solitons and those of the linear
QTI governed by Eqs. (\ref{new_LQTI_1})-(\ref{new_LQTI_4}) is significant,
indicating that these equations are not suitable for characterizing the
topological properties of the NLQTI. Instead, we will approach the
topological property of the NLQTI in the strongly nonlinear regime from a
different perspective.

From Eqs. (\ref{eq1})-(\ref{eq4}), as the onsite energies dominate the
difference between the intracell and intercell hopping strengths, we can
approximate Eqs. (\ref{eq1})-(\ref{eq4}) as follows:
\begin{eqnarray}
\mathrm{i}\frac{dV_{m,n}^{(1)}}{dT} &=&\delta _{1}V_{m,n}^{(1)}+\chi
_{C}V_{m,n}^{(3)}+\chi _{C}V_{m+1,n}^{(3)}+\chi _{C}V_{m,n}^{(4)}+\chi
_{C}V_{m,n+1}^{(4)}  \nonumber \\
&&+g\left( V_{m,n}^{(1)}\right) V_{m,n}^{(1)},  \label{NL_eq1} \\
\mathrm{i}\frac{dV_{m,n}^{(2)}}{dT} &=&\delta _{2}V_{m,n}^{(2)}-\chi
_{L}V_{m,n}^{(3)}-\chi _{L}V_{m,n-1}^{(3)}+\chi _{C}V_{m,n}^{(4)}+\chi
_{C}V_{m-1,n}^{(4)}  \nonumber \\
&&+g\left( V_{m,n}^{(2)}\right) V_{m,n}^{(2)},  \label{NL_eq2} \\
\mathrm{i}\frac{dV_{m,n}^{(3)}}{dT} &=&\delta _{3}V_{m,n}^{(3)}-\chi
_{L}V_{m,n}^{(2)}-\chi _{L}V_{m,n+1}^{(2)}+\chi _{C}V_{m,n}^{(1)}+\chi
_{C}V_{m-1,n}^{(1)}  \nonumber \\
&&+g\left( V_{m,n}^{(3)}\right) V_{m,n}^{(3)},  \label{NL_eq3} \\
\mathrm{i}\frac{dV_{m,n}^{(4)}}{dT} &=&\delta _{4}V_{m,n}^{(4)}+\chi
_{C}V_{m,n}^{(1)}+\chi _{C}V_{m,n-1}^{(1)}+\chi _{C}V_{m,n}^{(2)}+\chi
_{C}V_{m+1,n}^{(2)}  \nonumber \\
&&+g\left( V_{m,n}^{(4)}\right) V_{m,n}^{(4)},  \label{NL_eq4}
\end{eqnarray}%
where $\chi _{C}=\left( \gamma _{C}+\lambda _{C}\right) /2$ and $\chi
_{L}=\left( \gamma _{L}+\lambda _{L}\right) /2$. Since $\chi _{C}=\chi _{L}$
under our circuit parameters, these equations describe a nonlinear lattice
with equal intracell and intercell hopping strengths, resulting in no
dimerization. In the linear case with $g=0$, this lattice is gapless and
does not support topological corner states. However, under the action of the
nonlinearity, this lattice can support conventional self-trapped states,
commonly referred to as surface solitons \cite%
{RMP83-247,PR463-1,RPP75-086401}. In our study, we denote these self-trapped
states as corner solitons because they are localized at the lattice corners.
In Fig. \ref{fig_topo}(f), we illustrate the dependence of the
eigenfrequency of the corner soliton on the maximum voltage in the lattice,
calculated using Eqs. (\ref{NL_eq1})-(\ref{NL_eq4}), represented by the red
curve. For comparison, we include the curve for the corner soliton obtained
from Eqs. (\ref{eq1})-(\ref{eq4}), indicated by the blue curve. The corner
solitons from both types of lattice models exist within the semi-infinite
gap, indicating that they share the same physical origin: their existence is
entirely induced by nonlinearity, rendering them topologically trivial.

When the maximum voltage becomes sufficiently large, that is, in the
ultra-strongly nonlinear regime, the hopping strengths $\chi _{C}$ and $\chi
_{L}$ can be neglected compared to the onsite energies \cite{PRB104-235420}.
In this case, Eqs. (\ref{NL_eq1})-(\ref{NL_eq4}) further reduce to
\begin{equation}
\mathrm{i}\frac{dV_{2,2}^{(1)}}{dT}=\delta _{1}V_{2,2}^{(1)}+g\left(
V_{2,2}^{(1)}\right) V_{2,2}^{(1)},
\end{equation}%
which governs a nonlinear single lattice site. Solving this equation reveals
that the corner soliton is localized at a single corner site.

\subsection{The stability analysis\label{app_corner_stability}}

The stability of nonlinear states is their crucially important property, as
only stable states can be excited and observed experimentally. We
investigated the dynamical stability of the nonlinear topological corner
states and topologically trivial corner solitons by means of the standard
linear-stability analysis. To this end, perturbed solutions
\begin{equation}
V_{m,n}^{(j)}=e^{-\mathrm{i}\bar{\omega}T}\left( \psi
_{m,n}^{(j)}+\varepsilon _{m,n}^{(j)}e^{-\mathrm{i}\zeta T}+\mu
_{m,n}^{(j)\ast }e^{\mathrm{i}\zeta ^{\ast }T}\right) ,
\end{equation}%
with $j=1,2,3,4$, were substituted into Eqs.~(\ref{eq1})-(\ref{eq4}). Here $%
\psi _{m,n}^{(j)}$ is the unperturbed solution representing the nonlinear
corner states or corner solitons, $\varepsilon _{m,n}^{(i)}$ and $\mu
_{m,n}^{(i)}$ are infinitesimal amplitudes of the perturbations, which
correspond to eigenvalue $\zeta $, and $\bar{\omega}$ is the normalized
eigenfrequency defined as $\bar{\omega}=\omega /\omega _{n}$. The nonlinear
corner states or corner solitons are linearly stable if the imaginary part
of $\zeta $ (i.e., the growth rate) is positive. The linearization leads to
the following equations:
\begin{eqnarray}
&&-\bar{\omega}\varepsilon _{m,n}^{\left( 1\right) }+\delta _{1}\varepsilon
_{m,n}^{\left( 1\right) }+\gamma _{C}\varepsilon _{m,n}^{\left( 3\right)
}+\gamma _{C}\varepsilon _{m,n}^{\left( 4\right) }+\lambda _{C}\varepsilon
_{m+1,n}^{\left( 3\right) }+\lambda _{C}\varepsilon _{m,n+1}^{\left(
4\right) } \nonumber \\
&&+g \left(\psi _{m,n}^{(1)}\right)\varepsilon _{m,n}^{\left( 1\right) }
+g_{1}\left(\psi _{m,n}^{(1)}\right)\psi _{m,n}^{(1)2}\mu _{m,n}^{\left( 1\right)}
+g_{1}\left(\psi _{m,n}^{(1)}\right)\left\vert \psi _{m,n}^{(1)}\right\vert^{2}\varepsilon _{m,n}^{\left( 1\right) } \nonumber \\
&=&\zeta \varepsilon _{m,n}^{\left(1\right) }, \\
&&\bar{\omega}\mu _{m,n}^{\left( 1\right) }-\delta _{1}\mu _{m,n}^{\left(
1\right) }-\gamma _{C}\mu _{m,n}^{\left( 3\right) }-\gamma _{C}\mu
_{m,n}^{\left( 4\right) }-\lambda _{C}\mu _{m+1,n}^{\left( 3\right)
}-\lambda _{C}\mu _{m,n+1}^{\left( 4\right) } \nonumber \\
&&-g\left(\psi _{m,n}^{(1)}\right)\mu_{m,n}^{\left( 1\right) }
-g_{1}\left(\psi _{m,n}^{(1)}\right)\psi _{m,n}^{(1)\ast 2}\varepsilon _{m,n}^{\left(
1\right) }-g_{1}\left(\psi _{m,n}^{(1)}\right)\left\vert \psi _{m,n}^{(1)}\right\vert
^{2}\mu _{m,n}^{\left( 1\right) } \nonumber \\
&=&\zeta \mu _{m,n}^{\left( 1\right) }, \\
&&-\bar{\omega}\varepsilon _{m,n}^{\left( 2\right) }+\delta _{2}\varepsilon
_{m,n}^{\left( 2\right) }+\gamma _{C}\varepsilon _{m,n}^{\left( 4\right)
}-\gamma _{L}\varepsilon _{m,n}^{\left( 3\right) }+\lambda _{C}\varepsilon
_{m-1,n}^{\left( 4\right) }-\lambda _{L}\varepsilon _{m,n-1}^{\left(
3\right) } \nonumber \\
&&+g\left(\psi _{m,n}^{(2)}\right)\varepsilon _{m,n}^{\left( 2\right) }
+g_{1}\left(\psi _{m,n}^{(2)}\right)\psi _{m,n}^{(2)2}\mu _{m,n}^{\left( 2\right)
}+g_{1}\left(\psi _{m,n}^{(2)}\right)\left\vert \psi _{m,n}^{(2)}\right\vert
^{2}\varepsilon _{m,n}^{\left( 2\right) } \nonumber \\
&=&\zeta \varepsilon _{m,n}^{\left(2\right) }, \\
&&\bar{\omega}\mu _{m,n}^{\left( 2\right) }-\delta _{2}\mu _{m,n}^{\left(
2\right) }-\gamma _{C}\mu _{m,n}^{\left( 4\right) }+\gamma _{L}\mu
_{m,n}^{\left( 3\right) }-\lambda _{C}\mu _{m-1,n}^{\left( 4\right)
}+\lambda _{L}\mu _{m,n-1}^{\left( 3\right) } \nonumber \\
&&-g\left(\psi _{m,n}^{(2)}\right)\mu_{m,n}^{\left( 2\right) }
-g_{1}\left(\psi _{m,n}^{(2)}\right)\psi _{m,n}^{(2)\ast 2}\varepsilon _{m,n}^{\left(
2\right) }-g_{1}\left(\psi _{m,n}^{(2)}\right)\left\vert \psi _{m,n}^{(2)}\right\vert
^{2}\mu _{m,n}^{\left( 2\right) } \nonumber \\
&=&\zeta \mu _{m,n}^{\left( 2\right) }, \\
&&-\bar{\omega}\varepsilon _{m,n}^{\left( 3\right) }+\delta _{3}\varepsilon
_{m,n}^{\left( 3\right) }+\gamma _{C}\varepsilon _{m,n}^{\left( 1\right)
}-\gamma _{L}\varepsilon _{m,n}^{\left( 2\right) }+\lambda _{C}\varepsilon
_{m-1,n}^{\left( 1\right) }-\lambda _{L}\varepsilon _{m,n+1}^{\left(
2\right) } \nonumber \\
&&+g\left(\psi _{m,n}^{(3)}\right)\varepsilon _{m,n}^{\left( 3\right) }
+g_{1}\left(\psi _{m,n}^{(3)}\right)\psi _{m,n}^{(3)2}\mu _{m,n}^{\left( 3\right)
}+g_{1}\left(\psi _{m,n}^{(3)}\right)\left\vert \psi _{m,n}^{(3)}\right\vert
^{2}\varepsilon _{m,n}^{\left( 3\right) }\nonumber \\
&=&\zeta \varepsilon _{m,n}^{\left(3\right) }, \\
&&\bar{\omega}\mu _{m,n}^{\left( 3\right) }-\delta _{3}\mu _{m,n}^{\left(
3\right) }-\gamma _{C}\mu _{m,n}^{\left( 1\right) }+\gamma _{L}\mu
_{m,n}^{\left( 2\right) }-\lambda _{C}\mu _{m-1,n}^{\left( 1\right)
}+\lambda _{L}\mu _{m,n+1}^{\left( 2\right) } \nonumber \\
&&-g\left(\psi _{m,n}^{(3)}\right)\mu_{m,n}^{\left( 3\right) }
-g_{1}\left(\psi _{m,n}^{(3)}\right)\psi _{m,n}^{(3)\ast 2}\varepsilon _{m,n}^{\left(
3\right) }-g_{1}\left(\psi _{m,n}^{(3)}\right)\left\vert \psi _{m,n}^{(3)}\right\vert
^{2}\mu _{m,n}^{\left( 3\right) } \nonumber \\
&=&\zeta \mu _{m,n}^{\left( 3\right) }, \\
&&-\bar{\omega}\varepsilon _{m,n}^{\left( 4\right) }+\delta _{4}\varepsilon
_{m,n}^{\left( 4\right) }+\gamma _{C}\varepsilon _{m,n}^{\left( 2\right)
}+\gamma _{C}\varepsilon _{m,n}^{\left( 1\right) }+\lambda _{C}\varepsilon
_{m+1,n}^{\left( 2\right) }+\lambda _{C}\varepsilon _{m,n-1}^{\left(
1\right) } \nonumber \\
&&+g\left(\psi _{m,n}^{(4)}\right)\varepsilon _{m,n}^{\left( 4\right) }
+g_{1}\left(\psi _{m,n}^{(4)}\right)\psi _{m,n}^{(4)2}\mu _{m,n}^{\left( 4\right)
}+g_{1}\left(\psi _{m,n}^{(4)}\right)\left\vert \psi _{m,n}^{(4)}\right\vert
^{2}\varepsilon _{m,n}^{\left( 4\right) } \nonumber \\
&=&\zeta \varepsilon _{m,n}^{\left(4\right) }, \\
&&\bar{\omega}\mu _{m,n}^{\left( 4\right) }-\delta _{4}\mu _{m,n}^{\left(
4\right) }-\gamma _{C}\mu _{m,n}^{\left( 2\right) }-\gamma _{C}\mu
_{m,n}^{\left( 1\right) }-\lambda _{C}\mu _{m+1,n}^{\left( 2\right)
}-\lambda _{C}\mu _{m,n-1}^{\left( 1\right) } \nonumber \\
&&-g\left(\psi _{m,n}^{(4)}\right)\mu_{m,n}^{\left( 4\right) }
-g_{1}\left(\psi _{m,n}^{(4)}\right)\psi _{m,n}^{(4)\ast 2}\varepsilon _{m,n}^{\left(
4\right) }-g_{1}\left(\psi _{m,n}^{(4)}\right)\left\vert \psi _{m,n}^{(4)}\right\vert
^{2}\mu _{m,n}^{\left( 4\right) } \nonumber \\
&=&\zeta \mu _{m,n}^{\left( 4\right) },
\end{eqnarray}%
where%
\begin{equation}
g_{1}\left( {\psi _{m,n}^{\left( j\right) }}\right) =\frac{J }{4\left( C_{%
\text{g}}+J C_{\text{L}}\right) }\left[ \frac{MC_{\text{L}}}{\left(
1+\left\vert \psi _{m,n}^{\left( j\right) }/v_{0}\right\vert \right) ^{M+1}}%
\frac{1}{v_{0}\left\vert \psi _{m,n}^{\left( j\right) }\right\vert }\right]
\frac{{\omega _{0}}}{\omega _{n}}
\end{equation}%
with $j=1,2,3,4$.

\begin{figure*}[t]
\centering
\includegraphics[width=17.2cm]{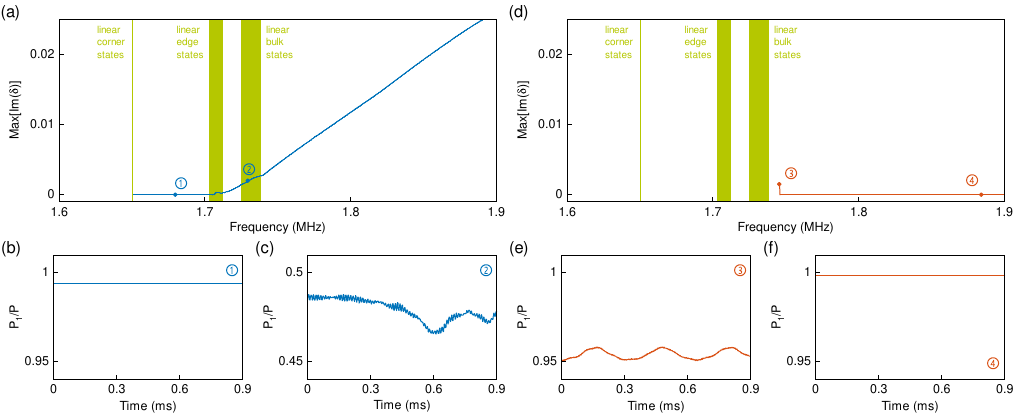}
\caption{\textbf{The stability analysis for the nonlinear topological corner
states and topologically trivial corner solitons.} (\textbf{a}) The maximum
perturbation growth rates, $\text{Max}\left[ \text{Im}\left( \protect\delta %
\right) \right] $, of the nonlinear corner states. (\textbf{b})-(\textbf{c})
The evolution of the eigenstates labeled in (a), with noise at the $\pm 2\%$
level added to the input. The ratio $P_{1}/P$ quantifies the fraction of the
total power carried by the sublattice of sites $1$ across the entire system.
(\textbf{d}) The maximum growth rates, $\text{Max}\left[ \text{Im}\left(
\protect\delta \right) \right] $, of the corner solitons. (\textbf{e})-(%
\textbf{f}) The evolution of the eigenstates labeled in (d), with the $\pm
2\%$ noise added to the input. }
\label{fig_stability_corner_state}
\end{figure*}

To confirm the expected stability or instability of the nonlinear corner
states and corner solitons, we simulated their temporal evolution in the
framework of full equations~(\ref{eq1})-(\ref{eq4}). For this purpose, we
used the Runge-Kutta algorithm, adding initial random perturbations at the
level of $\pm 2\%$ to the eigenstates and evaluating ratio $P_{1}/P$ during
the evolution, where $P_{j}=\sum\limits_{m,n}\left\vert \psi
_{m,n}^{(j)}\right\vert ^{2}$ ($j=1,2,3,4$) and $P=\sum%
\limits_{j=1}^{4}P_{j} $. This ratio quantifies the fraction of the power
carried by the sublattice $j$ across the entire lattice. A sufficiently
small time-matching step is chosen to ensure the accuracy of the simulation
results.

Figures \ref{fig_stability_corner_state}(a)-(c) illustrate the stability of
the nonlinear topological corner states. In panel (a), we observe that the
maximum perturbation growth rate, $\text{Max}\left[ \text{Im}\left( \delta
\right) \right] $, remains zero until the frequency enters the band
associated with the linear edge states. This indicates that the nonlinear
corner states are linearly stable when they are well localized at the
lattice corner. When the frequency enters the edge band, $\text{Max}\left[
\text{Im}\left( \delta \right) \right] $ gradually increases following the
increase of the frequency, leading to the linear instability of the
delocalized nonlinear corner states. The simulation results produced by the
direct simulations of the evolution confirm the predictions made by the
linear stability analysis. As shown in Fig. \ref{fig_stability_corner_state}%
(b), for state \textcircled{1}, which is a linearly stable one, ratio $%
P_{1}/P$ remains constant during the evolution. On the other hand, state %
\textcircled{2}, which belongs to the linearly unstable regime, $P_{1}/P$
exhibits significant variance, as seen in Fig. \ref%
{fig_stability_corner_state}(c).

Figures \ref{fig_stability_corner_state}(d)-(f) display results of the
stability analysis for the topologically trivial corner solitons. In panel
(d), the maximum perturbation growth rate, $\text{Max}\left[ \text{Im}\left(
\delta \right) \right] $, is nonzero only at the smallest frequency. At all
other frequencies, $\text{Max}\left[ \text{Im}\left( \delta \right) \right]
=0$, which implies that the corner solitons are linearly unstable at the
smallest frequency, and stable at all other frequencies. This conclusion is
confirmed by the simulations of the perturbed evolution. As shown in Fig. %
\ref{fig_stability_corner_state}(e), for state \textcircled{3}, which
corresponds to the smallest frequency, ratio $P_{1}/P$ slightly oscillates
during the evolution, demonstrating its weak instability. In contrast, for
state \textcircled{4}, which belongs to the linearly stable regime, $P_{1}/P$
remains constant, as seen in Fig. \ref{fig_stability_corner_state}(f).

\section{Bulk solitons\label{app_bulk}}

We investigated the solitons that reside in the bulk of the NLQTI. We
employed the anti-continuum limit to identify the initial solutions, which
were then used to produce the bulk soliton solutions. Subsequently, we
conducted the stability analysis for various types of bulk solitons. Note
that for a NLQTI, governed by Eqs. (\ref{eq1})-(\ref{eq4}), the parameters
satisfy the costraints $\delta _{1}=\delta _{2}=\delta _{3}=\delta
_{4}\equiv \delta $, $\lambda _{C}=\lambda _{L}\equiv \lambda $, and $\gamma
_{C}=\gamma _{L}\equiv \gamma $.

\subsection{Soliton solutions in the anti-continuum limit}

First, we set the intercell hopping to be $\lambda =0$, which simplifies the
governing equations to:
\begin{eqnarray}
\left[ \bar{\omega}-\delta -g\left( \psi _{m,n}^{(1)}\right) \right] \psi
_{m,n}^{\left( 1\right) } &=&\gamma \psi _{m,n}^{\left( 3\right) }+\gamma
\psi _{m,n}^{\left( 4\right) }, \\
\left[ \bar{\omega}-\delta -g\left( \psi _{m,n}^{(2)}\right) \right] \psi
_{m,n}^{\left( 2\right) } &=&\gamma \psi _{m,n}^{\left( 4\right) }-\gamma
\psi _{m,n}^{\left( 3\right) }, \\
\left[ \bar{\omega}-\delta -g\left( \psi _{m,n}^{(3)}\right) \right] \psi
_{m,n}^{\left( 3\right) } &=&\gamma \psi _{m,n}^{\left( 1\right) }-\gamma
\psi _{m,n}^{\left( 2\right) }, \\
\left[ \bar{\omega}-\delta -g\left( \psi _{m,n}^{(4)}\right) \right] \psi
_{m,n}^{\left( 4\right) } &=&\gamma \psi _{m,n}^{\left( 2\right) }+\gamma
\psi _{m,n}^{\left( 1\right) }.
\end{eqnarray}%
Solving these equations in the general form is quite challenging. Here, we
only consider four special cases.

(1) We assume that $\psi _{m,n}^{(2)}=0$ and $\psi _{m,n}^{(3)}=\psi
_{m,n}^{(4)}$. Under these conditions, the four equations reduce to two:
\begin{eqnarray}
\left[ \bar{\omega}-\delta -g\left(\psi _{m,n}^{(1)}\right)\right] \psi _{m,n}^{\left(
1\right) } &=&2\gamma \psi _{m,n}^{\left( 3\right) },  \label{AC_eq1} \\
\left[ \bar{\omega}-\delta -g\left(\psi _{m,n}^{(3)}\right)\right] \psi _{m,n}^{\left(
3\right) } &=&\gamma \psi _{m,n}^{\left( 1\right) }.  \label{AC_eq2}
\end{eqnarray}%
We also assume that the solutions are real. These equations can then be
solved numerically, usually supporting three types of soliton solutions. For
the sake of simplicity in comparison, we refer to these soliton solutions as
quasi-symmetric solitons, quasi-antisymmetric solitons, and asymmetric
solitons.

(2) We assume that $\psi _{m,n}^{(4)}=0$ and $\psi _{m,n}^{(1)}=-\psi
_{m,n}^{(2)}$. Under these assumptions, the four equations reduce to the
following pair:
\begin{eqnarray}
\left[ \bar{\omega}-\delta -g\left(\psi _{m,n}^{(1)}\right)\right] \psi _{m,n}^{\left(
1\right) } &=&\gamma \psi _{m,n}^{\left( 3\right) }, \\
\left[ \bar{\omega}-\delta -g\left(\psi _{m,n}^{(3)}\right)\right] \psi _{m,n}^{\left(
3\right) } &=&2\gamma \psi _{m,n}^{\left( 1\right) }.
\end{eqnarray}%
These equations can be obtained from those in case (1) by a transformation: $%
\psi _{m,n}^{\left( 1\right) }\rightarrow \psi _{m,n}^{\left( 3\right) }$, $%
\psi _{m,n}^{\left( 3\right) }\rightarrow \psi _{m,n}^{\left( 1\right) }$, $%
\psi _{m,n}^{\left( 2\right) }\rightarrow -\psi _{m,n}^{\left( 4\right) }$,
and $\psi _{m,n}^{\left( 4\right) }\rightarrow -\psi _{m,n}^{\left( 2\right)
}$. The solitons derived from this case are tantamount to those in case (1)
due to reflection symmetry $m_{x}$.

(3) We set $\psi _{m,n}^{(3)}=0$ and $\psi _{m,n}^{(1)}=\psi _{m,n}^{(2)}$.
Under these assumptions, the four equations reduce to the following pair:
\begin{eqnarray}
\left[ \bar{\omega}-\delta -g\left(\psi _{m,n}^{(2)}\right)\right] \psi _{m,n}^{\left(
2\right) } &=&\gamma \psi _{m,n}^{\left( 4\right) }, \\
\left[ \bar{\omega}-\delta -g\left(\psi _{m,n}^{(4)}\right)\right] \psi _{m,n}^{\left(
4\right) } &=&2\gamma \psi _{m,n}^{\left( 2\right) }.
\end{eqnarray}%
These equations can be obtained from those in case (1) by another set of
transformations: $\psi _{m,n}^{\left( 1\right) }\rightarrow \psi
_{m,n}^{\left( 4\right) }$, $\psi _{m,n}^{\left( 3\right) }\rightarrow \psi
_{m,n}^{\left( 2\right) }$, $\psi _{m,n}^{\left( 4\right) }\rightarrow \psi
_{m,n}^{\left( 1\right) }$, and $\psi _{m,n}^{\left( 2\right) }\rightarrow
\psi _{m,n}^{\left( 3\right) }$. In this case, the solitons are tantamount
in case (1) due to reflection symmetry $m_{y}$.

(4) We assume that $\psi _{m,n}^{(1)}=0$ and $\psi _{m,n}^{(3)}=-\psi
_{m,n}^{(4)}$. Under these assumptions, the four equations reduce to another
set of two equations:
\begin{eqnarray}
\left[ \bar{\omega}-\delta -g\left(\psi _{m,n}^{(2)}\right)\right] \psi _{m,n}^{\left(
2\right) } &=&2\gamma \psi _{m,n}^{\left( 4\right) }, \\
\left[ \bar{\omega}-\delta -g\left(\psi _{m,n}^{(4)}\right)\right] \psi _{m,n}^{\left(
4\right) } &=&\gamma \psi _{m,n}^{\left( 2\right) }.
\end{eqnarray}%
They can be obtained from those in case (2) by an appropriate
transformation, $\psi _{m,n}^{\left( 1\right) }\rightarrow \psi
_{m,n}^{\left( 4\right) }$, $\psi _{m,n}^{\left( 3\right) }\rightarrow \psi
_{m,n}^{\left( 2\right) }$, $\psi _{m,n}^{\left( 4\right) }\rightarrow \psi
_{m,n}^{\left( 1\right) }$, and $\psi _{m,n}^{\left( 2\right) }\rightarrow
\psi _{m,n}^{\left( 3\right) }$. The solitons derived from this case are
tantamount to those in case (2) due to reflection symmetry $m_{y}$.

\begin{figure}[t]
\centering
\includegraphics[width=8.6cm]{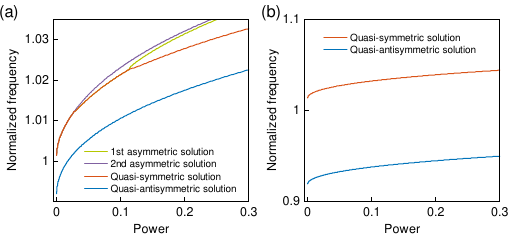}
\caption{\textbf{Soliton families obtained in the anti-continuum limit.} (a)
The solutions obtained in the limit of $\protect\lambda =0$. (b) The
solutions obtained in the limit of $\protect\gamma =0$.}
\label{fig_AC}
\end{figure}

\begin{figure*}[t]
\centering
\includegraphics[width=17.2cm]{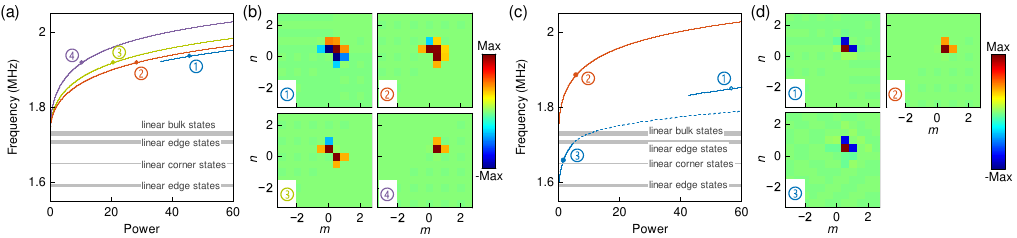}
\caption{\textbf{Bulk solitons in the NLQTI.} (\textbf{a})-(\textbf{b})
Bulk solitons that are the continuation of the solitons found at $%
\protect\lambda =0$. Panel (a) shows the relationship between the frequency
and soliton power, with panel (b) presenting voltage distributions of
typical solitons labeled in (a). The solid blue and red curves represent the
quasi-antisymmetric and quasi-symmetric bulk solitons, respectively, while
the yellow-green and purple curves correspond to the two types of asymmetric
bulk solitons. (\textbf{c})-(\textbf{d}) Bulk solitons that are
continuations of the solitons found at $\protect\gamma =0$. Panel (c)
presents the relationship between the frequency and soliton power, (d)
displaying voltage distributions of typical solitons labeled in (c).
The solid blue and red curves again denote the quasi-antisymmetric and
quasi-symmetric bulk solitons, respectively, while the dashed blue curve indicates
the delocalized states arising from one type of the quasi-antisymmetric bulk soliton.
It is worthy to note that there are two
branches of quasi-antisymmetric bulk solitons: one residing in the middle
finite gap between bands of the linear edge states, and the other one in the
semi-infinite gap.}
\label{fig_gap_solitons}
\end{figure*}

From the above discussion, one finds that the four cases are tantamount to
one another, and the soliton solutions in these cases can be transformed by
symmetry operations. Therefore, we now focus the study exclusively on the
soliton solutions in case (1). In the context of the circuit lattice, $%
\lambda =0$ implies $C_{2}=0$ and $L_{2}=\infty $. In the linear limit with $%
g=0$, Eqs. (\ref{AC_eq1}) and (\ref{AC_eq2}) yield only two solutions: one
is $\psi _{m,n}^{\left( 1\right) }=\sqrt{2}\psi _{m,n}^{\left( 3\right) }$
with the normalized eigenfrequency $\bar{\omega}=\delta +\sqrt{2}\gamma $,
and the other one is $\psi _{m,n}^{\left( 1\right) }=-\sqrt{2}\psi
_{m,n}^{\left( 3\right) }$ with $\bar{\omega}=\delta -\sqrt{2}\gamma $. In
the presence of the nonlinearity, i.e., when $g\neq 0$, Eqs. (\ref{AC_eq1})
and (\ref{AC_eq2}) may support additional solutions. Fig. \ref{fig_AC}(a)
presents the relationship between the soliton's power and normalized
frequency $\bar{\omega}$, with the power defined as $P=|\psi _{m,n}^{\left(
1\right) }|^{2}+|\psi _{m,n}^{\left( 3\right) }|^{2}$. From the figure, we
observe that Eqs. (\ref{AC_eq1}) and (\ref{AC_eq2}) support four types of
soliton solutions: quasi-symmetric, quasi-antisymmetric, and two asymmetric
ones. The quasi-symmetric and quasi-antisymmetric solutions (plotted by the
red and blue curves, respectively) bifurcate from the linear-limit
solutions, while the two asymmetric solutions (yellow-green and purple
curves) both bifurcate from the quasi-symmetric one.

Second, we set the intracell hopping to be $\gamma =0$, which reduces the
governing equations to the following system:
\begin{eqnarray}
\left[ \bar{\omega}-\delta -g\left(\psi _{m,n}^{(1)}\right)\right] \psi _{m,n}^{\left(
1\right) } &=&\lambda \psi _{m+1,n}^{\left( 3\right) }+\lambda \psi
_{m,n+1}^{\left( 4\right) }, \\
\left[ \bar{\omega}-\delta -g\left(\psi _{m+1,n+1}^{(2)}\right)\right] \psi
_{m+1,n+1}^{\left( 2\right) } &=&\lambda \psi _{m,n+1}^{\left( 4\right)
}-\lambda \psi _{m+1,n}^{\left( 3\right) }, \\
\left[ \bar{\omega}-\delta -g\left(\psi _{m+1,n}^{(3)}\right)\right] \psi
_{m+1,n}^{\left( 3\right) } &=&\lambda \psi _{m,n}^{\left( 1\right)
}-\lambda \psi _{m+1,n+1}^{\left( 2\right) }, \\
\left[ \bar{\omega}-\delta -g\left(\psi _{m,n+1}^{(4)}\right)\right] \psi
_{m,n+1}^{\left( 4\right) } &=&\lambda \psi _{m+1,n+1}^{\left( 2\right)
}+\lambda \psi _{m,n}^{\left( 1\right) }.
\end{eqnarray}%
This case too gives rise to four special solutions. Here, we focus on the
case with $\psi _{m+1,n+1}^{\left( 2\right) }=0$ and $\psi
_{m+1,n}^{(3)}=\psi _{m,n+1}^{\left( 4\right) }$, when the four equations
reduce to two:
\begin{eqnarray}
\left[ \bar{\omega}-\delta -g\left(\psi _{m,n}^{(1)}\right)\right] \psi _{m,n}^{\left(
1\right) } &=&2\lambda \psi _{m+1,n}^{\left( 3\right) },  \label{AC_eq3} \\
\left[ \bar{\omega}-\delta -g\left(\psi _{m+1,n}^{(3)}\right)\right] \psi
_{m+1,n}^{\left( 3\right) } &=&\lambda \psi _{m,n}^{\left( 1\right) }.
\label{AC_eq4}
\end{eqnarray}

In the context of the circuit lattice, $\gamma =0$ implies $C_{1}=0$ and $%
L_{1}=\infty $. In the linear limit with $g=0$, Eqs. (\ref{AC_eq3}) and (\ref%
{AC_eq4}) yield only two solutions: one with $\psi _{m,n}^{\left( 1\right) }=%
\sqrt{2}\psi _{m+1,n}^{\left( 3\right) }$ and the normalized eigenfrequency $%
\bar{\omega}=\delta +\sqrt{2}\lambda $, and the other one with $\psi
_{m,n}^{\left( 1\right) }=-\sqrt{2}\psi _{m+1,n}^{\left( 3\right) }$ and $%
\bar{\omega}=\delta -\sqrt{2}\lambda $. In contrast to the case of $\lambda
=0$, only two solutions exist if the nonlinearity is present: one
quasi-symmetric and one quasi-antisymmetric, as shown in Fig. \ref{fig_AC}%
(b). Both the quasi-symmetric and quasi-antisymmetric solutions bifurcate
from the linear-limit ones.

\subsection{The existence of bulk solitons\label{app_bulk_existence}}

\begin{figure}[t]
\centering
\includegraphics[width=8.6cm]{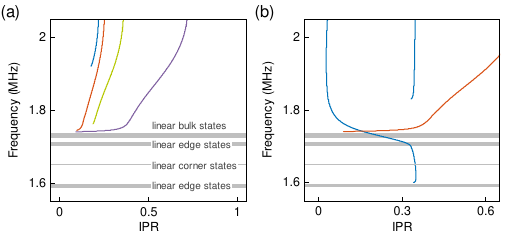}
\caption{\textbf{IPRs of the bulk solitons in the NLQTI.} (\textbf{a})
IPRs of the bulk solitons that are continuations of the solitons found at $%
\protect\lambda =0$. The blue and red curves represent the
quasi-antisymmetric and quasi-symmetric bulk solitons, respectively, while
the yellow-green and purple curves correspond to the two types of asymmetric
bulk solitons. (\textbf{b}) IPRs of the bulk solitons that are continuations
of the solitons found at $\protect\gamma =0$. The blue and red curves again
denote the quasi-antisymmetric and quasi-symmetric bulk solitons,
respectively. Note that the delocalized states arising from the quasi-antisymmetric bulk
solitons are also represented in the figure.}
\label{fig_bulk_IPR}
\end{figure}

First, extending the analysis, we investigate the bulk solitons that are
continuations of the solitons in the case of $\lambda =0$. We use the
soliton solutions obtained in the anti-continuum limit as the input in the
Newton's method and find the soliton solutions, gradually increasing $C_{2}$
from $0$ to the target value. Figure \ref{fig_gap_solitons}(a) presents the
relationship between the frequency and soliton power $P=\sum\limits_{m,n,j}%
\left\vert \psi _{m,n}^{(j)}\right\vert ^{2}$. The solid blue and red curves
represent the quasi-antisymmetric and quasi-symmetric bulk solitons,
respectively, while the solid yellow-green and purple curves correspond to two
types of asymmetric bulk solitons. Voltage distributions in typical solitons
labeled in (a) are shown in Fig. \ref{fig_gap_solitons}(b). From Fig. \ref%
{fig_gap_solitons}(a), we find that all four types of the bulk solitons
reside in the semi-infinite gap. These soliton branches terminate at small
powers, due to the inability of the Newton's method to capture poorly
localized solutions. All these bulk solitons are conventional lattice
solitons. As shown in Fig. \ref{fig_bulk_IPR}(a), they are more localized at
higher power, which is equivalent to enhanced nonlinearity.

Second, we investigate the bulk solitons that are continuations of the
soliton solutions found at $\gamma =0$. As there are only two solutions in
the anti-continuum limit, the bulk solitons in this case also exhibit solely
quasi-antisymmetric and quasi-symmetric distributions. However, we find that
there are two different branches of quasi-antisymmetric bulk
solitons: one residing in the middle finite gap between bands of the linear
edge states, and the other one in the semi-infinite gap, as shown in Fig. %
\ref{fig_gap_solitons}(c). The solid blue and red curves again denote the
quasi-antisymmetric and quasi-symmetric bulk solitons, respectively,
while the dashed blue curve indicates
the delocalized states arising from one type of the quasi-antisymmetric bulk soliton.
The voltage distributions of typical solitons labeled in (c)
are plotted in Fig. \ref{fig_gap_solitons}(d). Figure \ref{fig_bulk_IPR}(b)
displays IPRs of these bulk solitons. For the quasi-symmetric and
quasi-antisymmetric ones, located in the semi-infinite gap, IPR increases
with the frequency. In contrast, the
quasi-antisymmetric bulk solitons residing in the middle finite gap exhibit
weak localization near the upper and lower band edges of the linear edge
states.

\subsection{Stability/instability of the bulk solitons}

\begin{figure}[t]
\centering
\includegraphics[width=8.6cm]{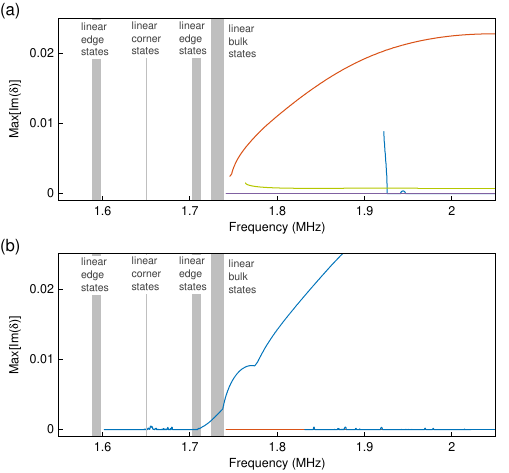}
\caption{\textbf{The linear stability analysis of the bulk solitons.} (%
\textbf{a}) The perturbation maximum growth rates, $\text{Max}\left[ \text{Im%
}\left( \protect\delta \right) \right] $, for the bulk solitons that are
continuations of the solitons at $\protect\lambda =0$. The blue and red
curves represent the quasi-antisymmetric and quasi-symmetric bulk solitons,
respectively, while the yellow-green and purple curves correspond to the two
types of asymmetric bulk solitons. (\textbf{b}) The maximum growth rates, $%
\text{Max}\left[ \text{Im}\left( \protect\delta \right) \right] $, of the
bulk solitons that are continuations of the solitons at $\protect\gamma =0$.
The blue and red curves denote the quasi-antisymmetric and quasi-symmetric
bulk solitons, respectively. It is important to note that there are two
branches of quasi-antisymmetric bulk solitons: one resides in the middle
finite gap, while the other is located in the semi-infinite gap.
Additionally, the delocalized states arising from the quasi-antisymmetric bulk
solitons are also represented in the figure.}
\label{fig_stability_bulk}
\end{figure}

We studied the stability of the bulk solitons by means of the
linear-stability analysis. Figure \ref{fig_stability_bulk}(a) shows the
maximum perturbation growth rates, $\text{Max}\left[ \text{Im}\left( \delta
\right) \right] $, of the bulk solitons that are continuations of the ones
found at $\lambda =0$. The blue and red curves represent the
quasi-antisymmetric and quasi-symmetric bulk solitons, respectively, while
the yellow-green and purple curves correspond to the two types of asymmetric
bulk solitons. The family of the quasi-antisymmetric solitons (denoted by the
blue curve) shows linear instability at low frequencies. In view of these
instabilities and the weak localization (see the blue curve in Fig. \ref%
{fig_bulk_IPR}(a)), the quasi-antisymmetric solitons can hardly be excited
using quench dynamics. Similarly, the quasi-symmetric solitons cannot be
excited either, due to the strong instability (see the red curve in Fig. \ref%
{fig_stability_bulk}(a)) and weak localization (the red curve in Fig. \ref%
{fig_bulk_IPR}(a)). The first type of the asymmetric bulk solitons (denoted
by the yellow-green curve in Fig. \ref{fig_stability_bulk}(a)), shows linear
instability too, which again denies the possibility to observe them in the
experiment. The asymmetric bulk solitons of the second type are linearly
stable (see the purple curve in Fig. \ref{fig_stability_bulk}(a)), their
localization is weak in the case of weak nonlinearity (see the purple curve
in Fig. \ref{fig_bulk_IPR}(a)). While the asymmetric bulk solitons of the
second type are strongly localized under the action of strong nonlinearity,
the IPR shows that their structure (voltage distribution) deviates from the
initial three-site excitation. Thus, considering these properties, all the
four types of the bulk solitons produced by the continuations of the
solitons at $\lambda =0$ are hardly observable.

Figure \ref{fig_stability_bulk}(b) illustrates the maximum growth rates, $%
\text{Max}\left[ \text{Im}\left( \delta \right) \right] $, of the bulk
solitons that are continuations of the solitons found at $\gamma =0$. The
blue and red curves represent the quasi-antisymmetric and quasi-symmetric
bulk solitons, respectively. Notably, there are two branches of
quasi-antisymmetric bulk solitons: one resides in the middle finite bandgap,
while the other is located in the semi-infinite gap.
Meanwhile, it is worth mentioning that the delocalized states arising from the
quasi-antisymmetric bulk solitons are also represented in the figure.
The quasi-antisymmetric bulk solitons residing in the middle finite bandgap are
generally linearly stable. Similarly, those located in the semi-infinite gap, while
exhibiting weak instabilities at several frequencies, are also generally stable,
particularly at frequencies near the linear bulk states. Given the
strong localizations of both types of quasi-antisymmetric bulk solitons and
the excellent correspondence between their state distributions and the
initial three-site excitations, these solitons are expected to be
experimentally observable, as supported by the experimental measurement
results presented in the main text. In contrast, while the quasi-symmetric
bulk solitons are linearly stable (see the red curve in Fig. \ref%
{fig_stability_bulk}(b)), their localization is weak under weak
nonlinearity, and their state distributions deviate from the initial
excitations under strong nonlinearity (see the red curve in Fig. \ref%
{fig_bulk_IPR}(b)). Consequently, the quasi-symmetric bulk solitons are
unlikely to be observed experimentally.

We have also conducted temporal evolutions with added noise to verify the
results obtained from the linear-stability analysis. The findings are in
good agreement with those presented in Fig. \ref{fig_stability_bulk}.

\section{The experimental implementation of the quench dynamics\label{app_exp}}

In this section, we provide a detailed description of our sample design and
fabrication, along with the experimental implementation of the quench
dynamics in the electric circuit lattice.

To minimize discrepancies between the experimental and theoretical results,
it is crucially important to minimize parasitic parameters and tolerance of
the circuit components. To achieve this, we utilized capacitors with low
equivalent series inductance (ESL) and a tolerance of $\pm 1\%$.
Additionally, we employed inductors featuring magnetic shielding and low
direct current resistance (DCR), carefully selecting components with the aid
of an inductance-capacitance-resistance (LCR) meter (HIOKI IM3536). The
tolerance for the inductance was also maintained at the level of $\pm 1\% $.
We fabricated the circuit lattice using standard PCB techniques, ensuring
that the inductors were sufficiently spaced to prevent mutual coupling. The
PCB traces were designed with a relatively large width of $0.75$ mm to
handle high currents, and the layouts were meticulously optimized to
minimize parasitic parameters and coupling with other components. To
facilitate the fabrication process, we divided the circuit board for a NLQTI
with $6\times 6$ unit cells into four sections, connecting them with
flexible flat cables (FFCs). For implementing the quench dynamics, we used
single-pole double-throw (SPDT) switches with two channels (ADG1636) to
control charging and discharging of capacitors and common-cathode diodes.
I-PEX MHF connectors were mounted on the PCB to serve as ports for
connecting external voltage sources.

\begin{figure}[t]
\centering
\includegraphics[width=7.9cm]{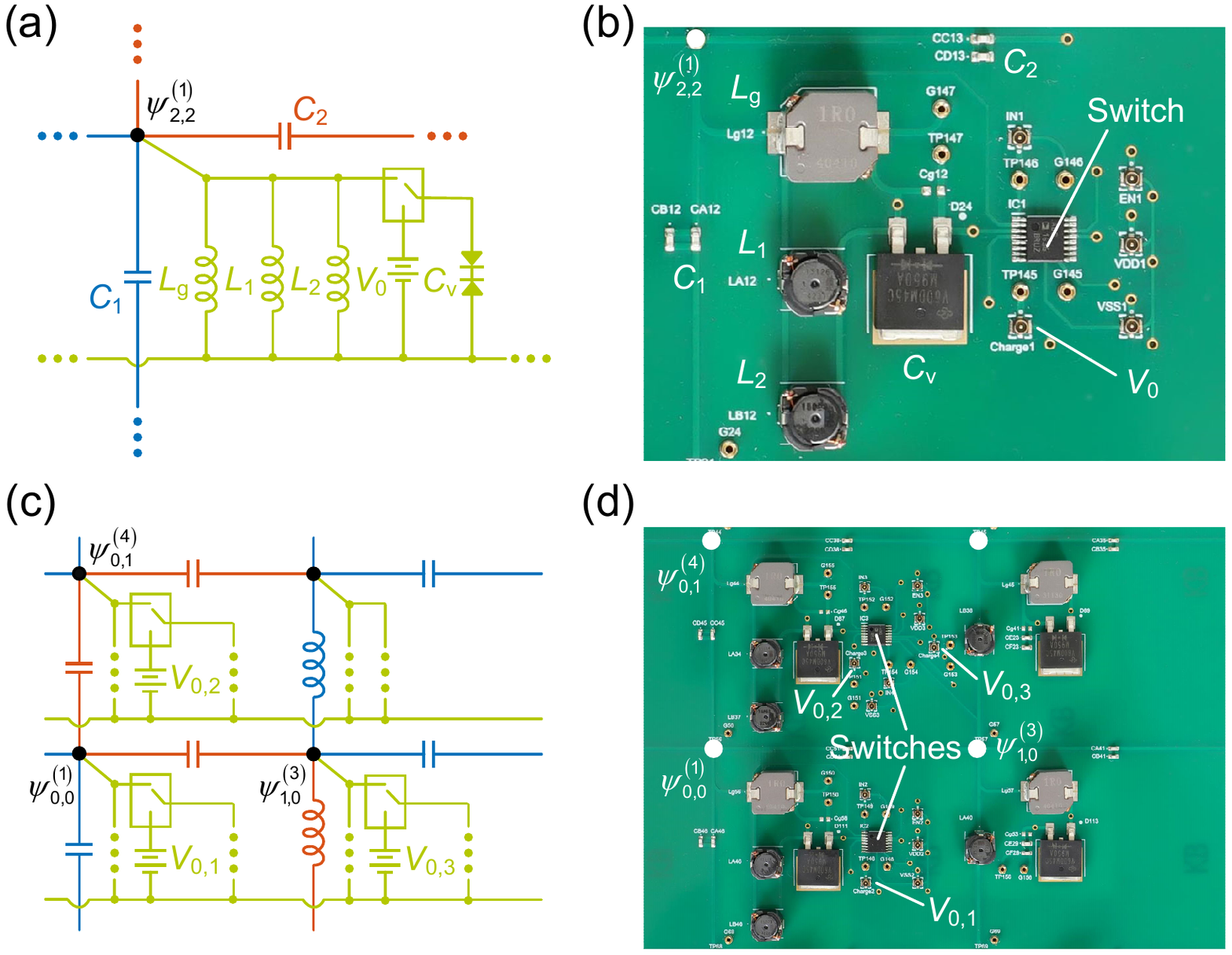}
\caption{\textbf{The experimental implementation of the NLQTI.} (\textbf{a})
A portion of the circuit diagram near the lattice corner. An SPDT switch is
used to control charging and discharging of the common-cathode diode $C_{%
\text{v}}$. (\textbf{b}) PCB near the lattice corner. The important circuit
components are labeled. (\textbf{c}) A portion of the circuit diagram near
the excitation sites in the bulk. Only the most important components are
labeled. (\textbf{d}) PCB near the excitation sites in the bulk.}
\label{fig_PCB}
\end{figure}

Figure \ref{fig_PCB}(a) displays a portion of the circuit diagram near the
lattice corner. At site $1$ of the $\left( 2,2\right) $-th unit cell, which
is the corner site of the circuit lattice, an SPDT switch controls the
charging and discharging of the common-cathode diode $C_{\text{v}}$. When
the SPDT switch is connected to the DC voltage source, diode $C_{\text{v}}$
is charged to constant voltage $V_{0}$, which is equal to $\psi _{0}$, as
defined in the main text. This charging operation corresponds to the
preparation of the initial state. Once the SPDT switch is toggled to the
circuit node, the charged diode discharges, and this discharging process
represents the temporal evolution of the initial voltage distribution
governed by Eqs. (\ref{eq1})--(\ref{eq4}). Figure \ref{fig_PCB}(b), with
important circuit components labeled in it, displays the PCB near the
lattice corner. The IPEX connector labeled \textquotedblleft Charge1" is
connected to a DC voltage source, providing voltage $V_{0}$. The IPEX
connector labeled \textquotedblleft IN1" is connected to an external digital
signal generated by an arbitrary-function generator (Tektronix AFG31022),
which is used to control the ON and OFF states of the SPDT switch. The
voltage signals at all circuit nodes were monitored using an oscilloscope
(Tektronix MDO34). This setup allows for the experimental observations of
the nonlinear topological corner states and topologically trivial corner
solitons.

Similarly, we implemented quench dynamics to verify the existence of the two
types of bulk solitons. Unlike the single-site excitations at the
lattice corner, in this case we utilized three-site excitations in the bulk
of the circuit lattice. Figure \ref{fig_PCB}(c) shows a portion of the
circuit diagram near the excitation sites in the bulk. For clarity's sake,
we label only the most important components in this diagram. Three SPDT
switches were employed to control charging and discharging at the three
lattice sites independently. Additionally, three different DC voltage
sources were used, allowing the phase of the initial state to be adjusted by
these voltage sources. Negative DC voltage is generated by inverting the
positive voltage, using a homemade inverting amplifier. Figure \ref{fig_PCB}%
(d) displays the PCB near the excitation sites in the bulk. In our
experiments, we utilized only two SPDT switches, as the ADG1636 switch
features two independent channels. The IPEX connectors labeled
\textquotedblleft Charge2\textquotedblright, \textquotedblleft
Charge3\textquotedblright, and \textquotedblleft Charge4\textquotedblright\
are connected to DC voltage sources providing voltages $V_{0,1}$, $V_{0,2}$,
and $V_{0,3}$, respectively. These voltages satisfy the conditions $V_{0,1}=-V_{0,3}=-V_{0,2}=\psi_{0}$.
To ensure the synchronization of the control
signals between the different switches, the IPEX connectors labeled
\textquotedblleft IN2\textquotedblright, \textquotedblleft
IN3\textquotedblright, and \textquotedblleft IN4\textquotedblright\ are
connected to the same external digital signal.

\begin{figure}[t]
\centering
\includegraphics[width=8.7cm]{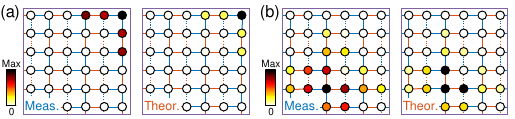}
\caption{\textbf{Voltage distributions in the medium-nonlinearity regime.}
(\textbf{a}) Voltage distributions when the corner site is excited with $\psi_{0}=0.5~\mathrm{V}$.
The left side shows the measurement result, and the right side shows the theoretical prediction.
(\textbf{b}) Voltage distributions when the three bulk sites are excited with $\psi_{0}=2.5~\mathrm{V}$.
Similarly, the left and right sides present the measurement result and the corresponding theoretical prediction, respectively.}
\label{fig_vol_medium}
\end{figure}

After conducting the measurements, we obtained voltage distributions at different time intervals for various excitation voltages $\psi_{0}$. The voltage envelopes were then derived from the temporal voltage signals. Taking circuit dissipation into account, we extracted the voltage distributions at $t = 9~\mu\mathrm{s}$ for corner site excitation and at $t = 8~\mu\mathrm{s}$ for the excitation of the three bulk sites. The selection of these time moments ensures accurate measurement of voltage signals with negligible background noise.

For the corner site excitation, the field distributions in the weakly nonlinear regime with $\psi_{0} = 0.1~\mathrm{V}$ and in the strongly nonlinear regime with $\psi_{0} = 2~\mathrm{V}$ are shown in Fig. \ref{fig3}(d), with the corresponding theoretical predictions displayed in Fig. \ref{fig3}(e). The experimentally measured and theoretically predicted field distributions in the moderately nonlinear regime with $\psi_{0} = 0.5~\mathrm{V}$ are presented in Fig. \ref{fig_vol_medium}(a).
By comparing the results, we find that in the weakly nonlinear regime, the voltage distribution is localized with sublattice polarization. In the moderately nonlinear regime, although there is no apparent delocalization, we observe that the sublattice polarization is suppressed due to hybridization with the edge modes. The delocalization behavior becomes more pronounced when considering the quench dynamics over a longer time evolution (see \ref{app_quench}). In the strongly nonlinear regime, the voltage distributions exhibit strong localization, with the sublattice polarization completely disappearing.

Similarly, for the excitation at the three bulk sites, the field distributions in the weakly nonlinear regime with $\psi_{0} = 0.7~\mathrm{V}$ and in the strongly nonlinear regime with $\psi_{0} = 7~\mathrm{V}$ are shown in Fig. \ref{fig4}(c), with the corresponding theoretical predictions provided in Fig. \ref{fig4}(d). The experimentally measured and theoretically predicted field distributions in the moderately nonlinear regime with $\psi_{0} = 2.5~\mathrm{V}$ are illustrated in Fig. \ref{fig_vol_medium}(b).
By comparing the results, we find that in both the weakly and strongly nonlinear regimes, the voltage distributions are localized. In contrast, in the moderately nonlinear regime, there is apparent delocalization. Thus, as nonlinearity increases, we observe a transition from localization to delocalization and back to localization. Furthermore, this transition becomes 
more pronounced when considering the quench dynamics over longer time evolution (see \ref{app_quench}).

\section{The quench dynamics in longer time evolution\label{app_quench}}

\begin{figure}[t]
\centering
\includegraphics[width=8.7cm]{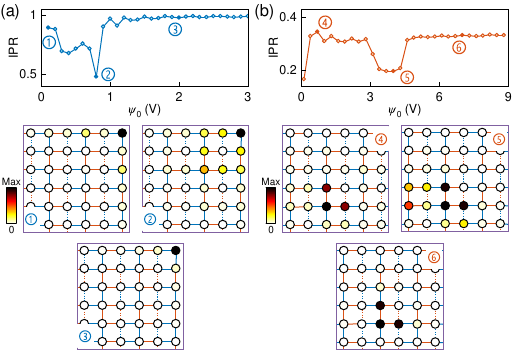}
\caption{\textbf{The quench dynamics, as displayed by the longer evolution.} (\textbf{a})-(%
\textbf{b}) IPRs of the voltage distributions for different initial
voltages, with insets displaying the voltage distributions at representative
values of $\protect\psi _{0}$. Panels (a) and (b) correspond to the
excitation applied at a single corner site and three bulk sites,
respectively. }
\label{fig_long}
\end{figure}

Due to the inherent circuit dissipation that limits experimental quench
dynamics, we have theoretically investigated the quench dynamics in the
course of longer evolution. Unlike the relatively short evolution observed
in Figs. \ref{fig3} and \ref{fig4} of the main text, which are limited to $t=9~\mathrm{\mu }%
\text{s}$ and $t=8~\mathrm{\mu }\text{s}$, respectively, we have here
examined the evolution extending up to $50~\mathrm{\mu }\text{s}$.

First, we investigated the case of excitations at a single corner site,
confirming the existence of persistent nonlinear topological corner states
and topologically trivial corner solitons. As shown in Fig. \ref{fig_long}%
(a), the IPRs of the voltage distributions for different initial voltages $%
\psi _{0}$ again exhibit the localization-delocalization-localization
transition, which is analyzed in detail in the main text.
The corresponding voltage distributions are localized for both small
and large initial voltages,
while they are evidently delocalized for medium initial voltages, as illustrated by the distributions at three representative
values of $\psi _{0}$ shown in the insets. The results produced by the
prolonged evolution are consistent with those presented in the main text.
This indicates that, for small initial voltages, the nonlinear corner states
are excited as long-lived ones by the quench dynamics. Similarly, long-lived
corner solitons are efficiently excited for large initial voltages. At
medium values of $\psi _{0}$, the voltage distributions become relatively
delocalized, as there are no localized states persisting in the
medium-nonlinearity regime. For small initial voltages, there is a slight
difference between the results of the extended and short-time evolution.
This discrepancy arises from the differences between the input distributions
and the eigenstates of the nonlinear corner states.

Second, we investigated the excitation at three bulk lattice sites,
reconfirming the existence of two types of the bulk solitons. As shown
in Fig. \ref{fig_long}(b), the IPRs of the voltage distributions for
different initial voltages $\psi _{0}$ definitely exhibit the
localization-delocaliation-localization transition. The voltage
distributions are localized for both small and large initial voltages,
while they are evidently delocalized for medium initial voltages, as illustrated by the distributions at three representative
values of $\psi _{0}$ shown in the insets.
For the first initial voltage, $\psi _{0}=0.1~\text{V}$, the IPR
is small because the bulk soliton demonstrates very weak localization in
the case of weak nonlinearity (see also \ref{app_bulk}). The
results produced by the extended evolution are again consistent with those
presented in the main text, indicating that, for small initial voltages, the
bulk solitons residing in the middle finite gap are definitely excited
by the quench dynamics. Similarly, the bulk solitons located in the
semi-infinite gap can also be effectively excited for large initial
voltages. At medium values of $\psi _{0}$, the voltage distributions become
relatively delocalized, as there are no localized states mainatined by
medium-strength nonlinearities. For small initial voltages, a discrepancy
exists between the results of the extended and short-time evolution. It
arises because the voltage distribution in the bulk solitons residing in
the middle finite bandgap cannot be accurately captured by the input
distribution, such as those with $\psi _{0,0}^{\left( 1\right) }=\psi _{0}$
and $\psi _{1,0}^{\left( 3\right) }=\psi _{0,1}^{\left( 4\right) }=-\psi
_{0} $.

\section{The summary of the work\label{summary}}

\begin{figure*}[t]
\centering
\includegraphics[width=12cm]{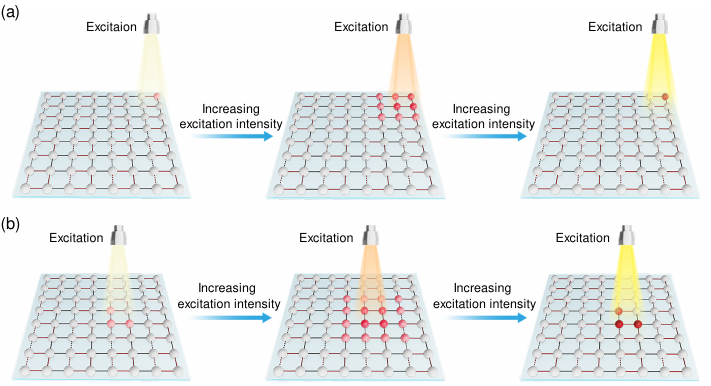}
\caption{\textbf{NLQTIs in the system with the onsite nonlinearity are
displayed by means of the field distributions corresponding to varying
excitation intensities.} (\textbf{a}) At a low intensity, the field
distribution is localized around the corner site, in form of nonlinear
topological corner states. As the excitation intensity increases, the field
spreads out, making the nonlinear corner states delocalized. At high
intensities, the field again gets tightly localized close to the corner
site, forming topologically trivial corner solitons. (\textbf{b}) When inner
sites (rather than the corner) are excited with different intensities, the
field distribution undergoes a localization-delocalization-localization
transition as well. The initial localized distribution corresponds to a bulk
soliton belonging to the middle finite gap in the case of weak
nonlinearity, while the final localized distribution corresponds to a bulk
soliton residing in the the semi-infinite gap, in the
strong-nonlinearity regime. }
\label{fig_illustration}
\end{figure*}

Based on the previous discussions, here we would like to summarize the
concept of the NLQTIs adopted in the present work and our main findings.

Building on the seminal lattice model for a linear QTI \cite{science357-61},
we have introduced onsite nonlinearity. In contrast to the QTI with
externally controlled hoppings \cite{PRB99-020304}, our system is
intrinsically nonlinear due to the self-interactions. As shown in Fig. \ref%
{fig_illustration}(a), when the corner site of the NLQTI is excited at low
intensity, the field distribution remains localized at the corner, forming
nonlinear topological corner states in the weakly nonlinear regime. As the
excitation intensity increases, the field distribution spreads out, as the
nonlinear corner states become delocalized in the regime of the moderately
nonlinearity strength (due to the fact that the respective solution branch
passes through a band of linear edge states). At high excitation
intensities, the field localizes again in the form of topologically trivial
corner solitons in the strongly nonlinear regime. A
localization-delocalization-localizion transition, similar to what we find
in the NLQTIs, was observed in a nonlinear Wannier-type HOTI \cite%
{nphys17-995}. Furthermore, we investigate the field distributions when
inner sites of the lattice (rather than the corner one) are locally excited
with varying intensities, as shown in Fig. \ref{fig_illustration}(b). In
this case, the field distribution again exhibits a
localized-delocalized-localized transition. The initial localized
distribution corresponds to a bulk soliton hosted by the middle finite
gap in the weakly nonlinear regime, while the final localized distribution
corresponds to a bulk soliton hosted by the semi-infinite gap in the
case of strong nonlinearity.

\section{Comparison with nonlinear Wannier-type higher-order topological
insulators\label{app_comparison}}

In this section, we compare our NLQTI with the previous experimental
realizations of nonlinear Wannier-type HOTIs. In 2019, Zangeneh-Nejad et al.
proposed the concept of nonlinear second-order TIs and demonstrated that
zero-dimensional corner states can be induced in a trivial insulator by
increasing the external pump intensity \cite{PRL123-053902}. The key to
their implementation lies in the nonlinear coupling coefficients of the
Wannier-type HOTI. Under weak nonlinearity, their structure is a trivial
insulator, lacking localized states in both the bulk and corner regions.
Under strong nonlinearity, their structure transitions to a Wannier-type
HOTI that supports localized corner states. In comparison, our NLQTI enables
nonlinearity-controlled switching between two types of localized states.
Under weak nonlinearity, our NLQTI supports nonlinear topological corner
states and bulk solitons residing in the middle finite gap. Under strong
nonlinearity, it supports topologically trivial corner solitons and bulk
solitons residing in the semi-infinite gap. This
localization-delocalization-localization transition enhances the active
manipulation of field distributions in HOTIs and may have applications in
photonics and cold atomic systems.

In 2021, Hu et al. \cite{LSA10-164} and Kirsch et al. \cite{nphys17-995}
independently realized nonlinear second-order photonic TIs. Both their
second-order TIs belong to the Wannier type, with nonlinearity introduced
through the onsite energies. Hu et al. utilized a 2D Su-Schrieffer-Heeger
(SSH) lattice that supports higher-order topological bound states in the
continuum (BICs) \cite{LSA10-164}. In this configuration, the
corner-localized states are embedded within the continuum of the bulk bands
rather than in the band gap \cite{PRB101-161116,PRL125-213901}. Under weak
nonlinearity, Hu et al. do not observe stationary beam dynamics; instead,
they find that the corner mode undergoes beating with the edge modes. In
contrast, our NLQTI supports stationary dynamics in both weakly and strongly
nonlinear regimes. The corner states of our QTI are stably separated from
other states, avoiding coupling between the corner states and edge states,
as long as the frequency is kept below the band of the linear edge states.

Kirsch et al. experimentally explored the nonlinear dynamics of light in
photonic HOTIs based on a kagome lattice \cite{nphys17-995}. Through the
observation of stationary beam dynamics, they identified the formation of
nonlinear topological corner states, as well as topologically trivial corner
solitons. A notable limitation of the kagome lattice is that the corner
states reside within the continuum of bulk modes when the lattice
dimerizations are weak, even though the topological invariants remain
non-vanishing. This limitation restricts the range of existence for
localized nonlinear corner states and imposes strict requirements on the
lattice parameters. In contrast, the corner states of our QTI are stably
separated from other states, even in weakly dimerized lattices, as long as
the quadrupole moment remains nonzero. This feature enables the realization
of well-localized nonlinear corner states and bulk solitons across a
broader range of frequencies, alleviating the stringent requirements on
lattice parameters.


\begin{thebibliography}{00}

\bibitem{RMP82-3045} M. Z. Hasan and C. L. Kane, Colloquium: Topological
insulators, Rev. Mod. Phys. 82, 3045 (2010).

\bibitem{RMP83-1057} X.-L. Qi and S.-C. Zhang, Topological insulators and
superconductors, Rev. Mod. Phys. 83, 1057 (2011).

\bibitem{science357-61} W. A. Benalcazar, B. A. Bernevig, and T. L. Hughes,
Quantized electric multipole insulators, Science 357, 61 (2017).

\bibitem{PRB96-245115} W. A. Benalcazar, B. A. Bernevig, and T. L. Hughes,
Electric multipole moments, topological multipole moment pumping, and chiral
hinge states in crystalline insulators, Phys. Rev. B 96, 245115 (2017).

\bibitem{nature555-342} M. Serra-Garcia, V. Peri, R. S\"{u}sstrunk, O. R.
Bilal, T. Larsen, L. G. Villanueva, and S. D. Huber, Observation of a
phononic quadrupole topological insulator, Nature 555, 342 (2018).

\bibitem{nature555-346} C. W. Peterson, W. A. Benalcazar, T. L. Hughes, and
G. Bahl, A quantized microwave quadrupole insulator with topologically
protected corner states, Nature 555, 346 (2018).

\bibitem{nphoton13-692} S. Mittal, V. V. Orre, G. Zhu, M. A. Gorlach, A.
Poddubny, and M. Hafezi, Photonic quadrupole topological phases, Nature
Photon. 13, 692 (2019).

\bibitem{nphys14-925} S. Imhof, C. Berger, F. Bayer, J. Brehm, L. W.
Molenkamp, T. Kiessling, F. Schindler, C. H. Lee, M. Greiter, T. Neupert,
and R. Thomale, Topolectrical-circuit realization of topological corner
modes, Nature Phys. 14, 925 (2018).

\bibitem{LSA9-1} S. Liu, S. Ma, Q. Zhang, L. Zhang, C. Yang, O. You, W. Gao,
Y. Xiang, T. J. Cui, and S. Zhang, Octupole corner state in a
three-dimensional topological circuit, Light: Science \& Applications 9, 1
(2020).

\bibitem{PRB100-201406} J. Bao, D. Zou, W. Zhang, W. He, H. Sun, and X.
Zhang, Topoelectrical circuit octupole insulator with topologically
protected corner states, Phys. Rev. B 100, 201406 (2019).

\bibitem{PRB102-100102} W. Zhang, D. Zou, J. Bao, W. He, Q. Pei, H. Sun, and
X. Zhang, Topolectrical-circuit realization of a four-dimensional
hexadecapole insulator, Phys. Rev. B 102, 100102 (2020).

\bibitem{ncommun11-2108} X. Ni, M. Li, M. Weiner, A. Al\`{u}, and A. B.
Khanikaev, Demonstration of a quantized acoustic octupole topological
insulator, Nat. Commun. 11, 2108 (2020).

\bibitem{PRL124-206601} Y. Qi, C. Qiu, M. Xiao, H. He, M. Ke, and Z. Liu,
Acoustic Realization of Quadrupole Topological Insulators, Phys. Rev. Lett.
124, 206601 (2020).

\bibitem{ncommun11-2442} H. Xue, Y. Ge, H.-X. Sun, Q. Wang, D. Jia, Y.-J.
Guan, S.-Q. Yuan, Y. Chong, and B. Zhang, Observation of an acoustic
octupole topological insulator, Nat. Commun. 11, 2442 (2020).

\bibitem{ncommun11-65} X. Zhang, Z.-K. Lin, H.-X. Wang, Z. Xiong, Y. Tian,
M.-H. Lu, Y.-F. Chen, and J.-H. Jiang, Symmetry-protected hierarchy of
anomalous multipole topological band gaps in nonsymmorphic metacrystals,
Nat. Commun. 11, 65 (2020).

\bibitem{staggered1} D. Cai, A. R. Bishop, and N. Gronbech-Jensen, Localized
states in discrete nonlinear Schr\"{o}dinger equation, Phys. Rev. Lett. 72,
591-595 (1994).

\bibitem{staggered2} Y. Lahini, A. Avidan, F. Pozzi, M. Sorel, R.
Morandotti, D. N. Christodoulides, and I. Silberberg, Anderson localization
and nonlinearity in one-dimensional disordered photonic lattices, Phys. Rev.
Lett. 100, 013906 (2008).

\bibitem{PRL120-026801} M. Ezawa, Higher-Order Topological Insulators and
Semimetals on the Breathing Kagome and Pyrochlore Lattices, Phys. Rev. Lett.
120, 026801 (2018).

\bibitem{PRB98-045125} M. Ezawa, Minimal models for Wannier-type
higher-order topological insulators and phosphorene, Phys. Rev. B 98, 045125
(2018).

\bibitem{nmater18-113} X. Ni, M. Weiner, A. Al\`{u}, and A. B. Khanikaev,
Observation of higher-order topological acoustic states protected by
generalized chiral symmetry, Nature Mater. 18, 113 (2019).

\bibitem{nmater18-108} H. Xue, Y. Yang, F. Gao, Y. Chong, and B. Zhang,
Acoustic higher-order topological insulator on a kagome lattice, Nature
Mater. 18, 108 (2019).

\bibitem{nphoton12-408} J. Noh, W. A. Benalcazar, S. Huang, M. J. Collins,
K. P. Chen, T. L. Hughes, and M. C. Rechtsman, Topological protection of
photonic mid-gap defect modes, Nature Photon. 12, 408 (2018).

\bibitem{nphys15-582} X. Zhang, H.-X. Wang, Z.-K. Lin, Y. Tian, B. Xie,
M.-H. Lu, Y.-F. Chen, and J.-H. Jiang, Second-order topology and
multidimensional topological transitions in sonic crystals, Nat. Phys. 15,
582 (2019).

\bibitem{PRL122-233903} B.-Y. Xie, G.-X. Su, H.-F. Wang, H. Su, X.-P. Shen,
P. Zhan, M.-H. Lu, Z.-L. Wang, and Y.-F. Chen, Visualization of Higher-Order
Topological Insulating Phases in Two-Dimensional Dielectric Photonic
Crystals, Phys. Rev. Lett. 122, 233903 (2019).

\bibitem{nmat18-1292} S. N. Kempkes, M. R. Slot, J. J. van den Broeke, P.
Capiod, W. A. Benalcazar, D. Vanmaekelbergh, D. Bercioux, I. Swart, and C.
Morais Smith, Robust zero-energy modes in an electronic higher-order
topological insulator, Nat. Mater. 18, 1292 (2019).

\bibitem{PRL122-204301} H. Fan, B. Xia, L. Tong, S. Zheng, and D. Yu,
Elastic Higher-Order Topological Insulator with Topologically Protected
Corner States, Phys. Rev. Lett. 122, 204301 (2019).

\bibitem{PRL122-233902} X.-D. Chen, W.-M. Deng, F.-L. Shi, F.-L. Zhao, M.
Chen, and J.-W. Dong, Direct Observation of Corner States in Second-Order
Topological Photonic Crystal Slabs, Phys. Rev. Lett. 122, 233902 (2019).

\bibitem{nphoton13-697} A. El Hassan, F. K. Kunst, A. Moritz, G. Andler, E.
J. Bergholtz, and M. Bourennane, Corner states of light in photonic
waveguides, Nature Photon. 13, 697 (2019).

\bibitem{sciadv6-eaay4166} M. Weiner, X. Ni, M. Li, A. Al\`{u}, and A. B.
Khanikaev, Demonstration of a third-order hierarchy of topological states in
a three-dimensional acoustic metamaterial, Sci. Adv. 6, eaay4166 (2020).

\bibitem{ncommun10-5331} X. Zhang, B.-Y. Xie, H.-F. Wang, X. Xu, Y. Tian,
J.-H. Jiang, M.-H. Lu, and Y.-F. Chen, Dimensional hierarchy of higher-order
topology in three-dimensional sonic crystals, Nat. Commun. 10, 5331 (2019).

\bibitem{ncommun16-3122} Z. Wang, Y. Meng, B. Yan, D. Zhao, L. Yang, J.
Chen, M. Cheng, T. Xiao, P. P. Shum, G.-G. Liu, Y. Yang, H. Chen, X. Xi, Z.
Zhu, B. Xie, and Z. Gao, Realization of a three-dimensional photonic
higher-order topological insulator, Nat. Commun. 16, 3122 (2025).

\bibitem{APR7-021306} D. Smirnova, D. Leykam, Y. Chong, and Y. Kivshar,
Nonlinear topological photonics, Appl. Phys. Rev. 7, 021306 (2020).

\bibitem{nphys20-905} A. Szameit and M. C. Rechtsman, Discrete nonlinear
topological photonics, Nature Phys. 20, 905 (2024).

\bibitem{arxiv-lee-1} H. Sahin, H. Akg\"{u}n, Z. B. Siu, S. M.
Rafi-Ul-Islam, J. F. Kong, M. B. A. Jalil, and C. H. Lee, Protected Chaos in
a Topological Lattice, Adv. Sci. 12, e03216 (2025).

\bibitem{arxiv-lee-2} H. Sahin, M. B. A. Jalil, and C. H. Lee, Topolectrical
Circuits $-$ Recent Experimental Advances and Developments, APL Electronic
Devices 1, 021503 (2025).

\bibitem{PR1093-1} H. Yang, L. Song, Y. Cao, and P. Yan, Circuit realization
of topological physics, Phys. Rep. 1093, 1 (2024).

\bibitem{ISAN} H. Zhong, V. O.Kompanets,Y. Zhang,Y. V. Kartashov, M. Cao, Y.
Li, S. A. Zhuravitskii, N. N. Skryabin, I. V. Dyakonov, A. A. Kalinkin, S.
P. Kulik, S. V. Chekalin, and V. N. Zadkov, Observation of nonlinear fractal
higher order topological insulator, Light Sci. Appl. 13, 264 (2024).

\bibitem{comment} B. A. Malomed, Prediction and observation of topological
modes in fractal nonlinear optics, Light Sci. Appl. 14, 29 (2025).

\bibitem{AM37-2500556} V. O. Kompanets, S. Feng, Y. Zhang, Y. V. Kartashov,
Y. Li, S. A. Zhuravitskii, N. N. Skryabin, A. V. Kireev, I. V. Dyakonov, A.
A. Kalinkin, C. Shang, S. P. Kulik, S. V. Chekalin, and V. N. Zadkov,
Observation of Nonlinear Topological Corner States Originating from
Different Spectral Charges, Adv. Mater. 37, 2500556 (2025).

\bibitem{CP8-451} C. Huang, A. V. Kireev, Y. Jiang, V. O. Kompanets, C. Shang, Y. V. Kartashov, S. A. Zhuravitskii, N. N. Skryabin, I. V. Dyakonov, A. A. Kalinkin, S. P. Kulik, F. Ye, and V. N. Zadkov, Observation of nonlinear higher-order topological insulators with unconventional boundary truncations, Commun. Phys. 8, 451 (2025).


\bibitem{PRA90-023813} M. J. Ablowitz, C. W. Curtis, and Y.-P. Ma, Linear
and nonlinear traveling edge waves in optical honeycomb lattices, Phys. Rev.
A 90, 023813 (2014).

\bibitem{PRL117-143901} D. Leykam and Y. D. Chong, Edge Solitons in
Nonlinear-Photonic Topological Insulators, Phys. Rev. Lett. 117, 143901
(2016).

\bibitem{PRA94-021801} Y. Lumer, M. C. Rechtsman, Y. Plotnik, and M. Segev,
Instability of bosonic topological edge states in the presence of
interactions, Phys. Rev. A 94, 021801 (2016).

\bibitem{PRL128-093901} Y. V. Kartashov, A. A. Arkhipova, S. A.
Zhuravitskii, N. N. Skryabin, I. V. Dyakonov, A. A. Kalinkin, S. P. Kulik,
V. O. Kompanets, S. V. Chekalin, L. Torner, and V. N. Zadkov, Observation of
Edge Solitons in Topological Trimer Arrays, Phys. Rev. Lett. 128, 093901
(2022).

\bibitem{ncommun15-9642} S. D. Hashemi and S. Mittal, Floquet topological
dissipative Kerr solitons and incommensurate frequency combs, Nat. Commun.
15, 9642 (2024).

\bibitem{PRX11-041057} S. Mukherjee and M. C. Rechtsman, Observation of
Unidirectional Solitonlike Edge States in Nonlinear Floquet Topological
Insulators, Phys. Rev. X 11, 041057 (2021).

\bibitem{PRL121-163901} D. A. Dobrykh, A. V. Yulin, A. P. Slobozhanyuk, A.
N. Poddubny, and Yu. S. Kivshar, Nonlinear Control of Electromagnetic
Topological Edge States, Phys. Rev. Lett. 121, 163901 (2018).

\bibitem{LSA9-147} Z. Hu, D. Bongiovanni, D. Juki\'{c}, E. Jajti\'{c}, S.
Xia, D. Song, J. Xu, R. Morandotti, H. Buljan, and Z. Chen, Nontrivial
coupling of light into a defect: the interplay of nonlinearity and topology,
Light Sci. Appl. 9, 147 (2020).

\bibitem{PRB102-115411} T. Tuloup, R. W. Bomantara, C. H. Lee, and J. Gong,
Nonlinearity induced topological physics in momentum space and real space,
Phys. Rev. B 102, 115411 (2020).

\bibitem{PRL133-116602} K. Bai, J.-Z. Li, T.-R. Liu, L. Fang, D. Wan, and M.
Xiao, Arbitrarily Configurable Nonlinear Topological Modes, Phys. Rev. Lett.
133, 116602 (2024).

\bibitem{ncommun16-422} K. Sone, M. Ezawa, Z. Gong, T. Sawada, N. Yoshioka,
and T. Sagawa, Transition from the topological to the chaotic in the
nonlinear Su-Schrieffer-Heeger model, Nat. Commun. 16, 422 (2025).

\bibitem{ncommun11-1902} Z. Zhang, R. Wang, Y. Zhang, Y. V. Kartashov, F.
Li, H. Zhong, H. Guan, K. Gao, F. Li, Y. Zhang, and M. Xiao, Observation of
edge solitons in photonic graphene, Nat. Commun. 11, 1902 (2020).

\bibitem{nphys17-1169} S. Mittal, G. Moille, K. Srinivasan, Y. K. Chembo,
and M. Hafezi, Topological frequency combs and nested temporal solitons,
Nat. Phys. 17, 1169 (2021).

\bibitem{science384-1356} C. J. Flower, M. Jalali Mehrabad, L. Xu, G.
Moille, D. G. Suarez-Forero, O. \"{O}rsel, G. Bahl, Y. Chembo, K.
Srinivasan, S. Mittal, M. Hafezi, Observation of topological frequency
combs, Science 384, 1356 (2024).

\bibitem{arxiv} R. Li, W.Wang, X. Kong, C. Shang, Y. Jia, G.-G. Liu, Y.
Liu, and B. Zhang, Self-Induced Topological Edge States
in a Lattice with Onsite Nonlinearity, arXiv:2504.11964 (2025).

\bibitem{PRL111-243905} Y. Lumer, Y. Plotnik, M. C. Rechtsman, and M. Segev,
Self-Localized States in Photonic Topological Insulators, Phys. Rev. Lett.
111, 243905 (2013).

\bibitem{science368-856} S. Mukherjee and M. C. Rechtsman, Observation of
Floquet solitons in a topological bandgap, Science 368, 856 (2020).

\bibitem{optica10-1310} S. Mukherjee and M. C. Rechtsman, Period-doubled
Floquet solitons, Optica 10, 1310 (2023).

\bibitem{PRL129-135501} G. Liu, J. Noh, J. Zhao, and G. Bahl, Self-Induced
Dirac Boundary State and Digitization in a Nonlinear Resonator Chain, Phys.
Rev. Lett. 129, 135501 (2022).

\bibitem{nphys18-678} N. Pernet, P. St-Jean, D. D. Solnyshkov, G. Malpuech,
N. C. Zambon, Q. Fontaine, B. Real, O. Jamadi, A. Lema\^{\i}tre, M. Morassi,
L. L. Gratiet, T. Baptiste, A. Harouri, I. Sagnes, A. Amo, S. Ravets, and J.
Bloch, Gap solitons in a one-dimensional driven-dissipative topological
lattice, Nat. Phys. 18, 678 (2022).

\bibitem{PRL118-023901} D. D. Solnyshkov, O. Bleu, B. Teklu, and G.
Malpuech, Chirality of Topological Gap Solitons in Bosonic Dimer Chains,
Phys. Rev. Lett. 118, 023901 (2017).

\bibitem{LPR13-1900223} D. A. Smirnova, L. A. Smirnov, D. Leykam, and Y. S.
Kivshar, Topological Edge States and Gap Solitons in the Nonlinear Dirac
Model, Laser \& Photonics Reviews 13, 1900223 (2019).

\bibitem{CP5-275} R. Li, X. Kong, D. Hang, G. Li, H. Hu, H. Zhou, Y. Jia, P.
Li, and Y. Liu, Topological bulk solitons in a nonlinear photonic Chern
insulator, Commun. Phys. 5, 275 (2022).

\bibitem{PRB105-L201111} R. Li, P. Li, Y. Jia, and Y. Liu, Self-localized
topological states in three dimensions, Phys. Rev. B 105, L201111 (2022).

\bibitem{NJP22-103058} Y.-L. Tao, N. Dai, Y.-B. Yang, Q.-B. Zeng, and Y. Xu,
Hinge solitons in three-dimensional second-order topological insulators, New
J. Phys. 22, 103058 (2020).

\bibitem{OL45-4710} Y. Zhang, Y. V. Kartashov, L. Torner, Y. Li, and A.
Ferrando, Nonlinear higher-order polariton topological insulator, Opt. Lett.
45, 4710 (2020).

\bibitem{PRB104-235420} M. Ezawa, Nonlinearity-induced transition in the
nonlinear Su-Schrieffer-Heeger model and a nonlinear higher-order
topological system, Phys. Rev. B 104, 235420 (2021).

\bibitem{PRA107-033514} S. K. Ivanov, Y. V. Kartashov, and L. Torner, Light
bullets in Su-Schrieffer-Heeger photonic topological insulators, Phys. Rev.
A 107, 033514 (2023).

\bibitem{CSF185-115188} Y. V. Kartashov, Solitons in higher-order
topological insulator created by unit cell twisting, Chaos, Solitons \&
Fractals 185, 115188 (2024).

\bibitem{PRB110-104307} K. Prabith, G. Theocharis, and R. Chaunsali,
Nonlinear corner states in a topologically nontrivial kagome lattice, Phys.
Rev. B 110, 104307 (2024).

\bibitem{NJP26-063004} J. Yi and C. Q. Chen, Delocalization and higher-order
topology in a nonlinear elastic lattice, New J. Phys. 26, 063004 (2024).

\bibitem{PRL123-053902} F. Zangeneh-Nejad and R. Fleury, Nonlinear
Second-Order Topological Insulators, Phys. Rev. Lett. 123, 053902 (2019).

\bibitem{LSA10-164} Z. Hu, D. Bongiovanni, D. Juki\'{c}, E. Jajti\'{c}, S.
Xia, D. Song, J. Xu, R. Morandotti, H. Buljan, and Z. Chen, Nonlinear
control of photonic higher-order topological bound states in the continuum,
Light Sci. Appl. 10, 164 (2021).

\bibitem{nphys17-995} M. S. Kirsch, Y. Zhang, M. Kremer, L. J. Maczewsky, S.
K. Ivanov, Y. V. Kartashov, L. Torner, D. Bauer, A. Szameit, and M.
Heinrich, Nonlinear second-order photonic topological insulators, Nat. Phys.
17, 995 (2021).

\bibitem{PRB99-020304} M. Serra-Garcia, R. S\"{u}sstrunk, and S. D. Huber,
Observation of quadrupole transitions and edge mode topology in an LC
circuit network, Phys. Rev. B 99, 020304 (2019).

\bibitem{RMP83-247} Y. V. Kartashov, B. A. Malomed, and L. Torner, Solitons
in nonlinear lattices, Rev. Mod. Phys. 83, 247 (2011).

\bibitem{PR463-1} F. Lederer, G. I. Stegeman, D. N. Christodoulides, G.
Assanto, M. Segev, and Y. Silberberg, Discrete solitons in optics, Phys.
Rep. 463, 1 (2008).

\bibitem{RPP75-086401} Z. Chen, M. Segev, and D. N. Christodoulides, Optical
spatial solitons: historical overview and recent advances, Rep. Prog. Phys.
75, 086401 (2012).

\bibitem{QF4-9} L. Wang, Z. Yan, Y. Zhu, and J. Zeng, Gap solitons and
vortices in two-dimensional spin-orbit-coupled Bose-Einstein condensates
loaded onto moir\'{e} optical lattices, Quantum. Front. 4, 9 (2025).

\bibitem{SP18-208} R. Liquito, M. Gon\c{c}alves, and E. Castro,
Quasiperiodic quadrupole insulators, SciPost Phys. 18, 208 (2025).

\bibitem{CP8-342} R. Li, X. Kong, W. Wang, Y. Wang, Y. Jia, H. Tao, P. Li,
Y. Liu, and B. A. Malomed, Observation of edge solitons and transitions
between them in a trimer circuit lattice, Commun. Phys. 8, 342 (2025).

\bibitem{Aubry} S. Aubry, Breathers in nonlinear lattices: Existence, linear
stability and quantization, Physica D 103, 201 (1997).

\bibitem{PR129-959} H. A. Gersch and G. C. Knollman, Quantum Cell Model for
Bosons, Phys. Rev. 129, 959 (1963).

\bibitem{nelectron1-178} Y. Hadad, J. C. Soric, A. B. Khanikaev, and A. Al%
\`{u}, Self-Induced Topological Protection in Nonlinear Circuit Arrays, Nat.
Electron. 1, 178 (2018).

\bibitem{PRB101-161116} W. A. Benalcazar and A. Cerjan, Bound states in the
continuum of higher-order topological insulators, Phys. Rev. B 101, 161116
(2020).

\bibitem{PRL125-213901} A. Cerjan, M. J\"{u}rgensen, W. A. Benalcazar, S.
Mukherjee, and M. C. Rechtsman, Observation of a Higher-Order Topological
Bound State in the Continuum, Phys. Rev. Lett. 125, 213901 (2020).

\end{thebibliography}
\end{document}